\documentclass{article} 
\usepackage{psfig} 
\usepackage{a4wide}  
\usepackage{verbatim}

\begin{document}

\title{Mesoscopic superconductors in the London limit: equilibrium properties  
and metastability} 
 
\author{E.~Akkermans$^1$, D.M.~Gangardt$^1$  and K.~Mallick$^2$  
 \\$^1$Department of Physics,  
 Technion, 32000 Haifa, Israel \\  $^2$Service de Physique Th\'eorique,  
 CEA Saclay, 91191 Gif-Sur-Yvette, France}

\maketitle

\begin{abstract} 
We present a study of the behaviour of metastable vortex states in 
 mesoscopic superconductors. Our analysis relies on 
 the London limit within which  
it is possible to derive closed analytical expressions for the magnetic field  
and the Gibbs free energy.  
We consider in particular the situation where 
 the vortices are symmetrically distributed along a closed ring.  
There, we obtain expressions for the confining Bean-Livingston barrier and for       
the magnetization  
which turns out to be paramagnetic away from thermodynamic equilibrium. At low  
temperature,  
the barrier is high enough for this regime to be observable. 
 We propose also a local description of  
both thermodynamic and metastable states based on elementary topological 
 considerations; we find structural phase transitions of  vortex patterns 
 between these metastable states and we calculate the corresponding 
 critical fields.   
\end{abstract} 
  

\section {Introduction} 
 
A significant amount of work has been recently devoted 
 to the study  of   aluminum  
superconducting  disks  in a  mesoscopic regime, in which the size of 
the sample is  comparable with both the  
coherence length $\xi$ and the London penetration depth $\lambda$.    
In a  first set of experiments \cite{geim1}  
the magnetization of such systems was measured at  
 thermodynamic equilibrium. Beyond the Meissner state, 
 a series of discontinuous jumps  appear when  
  the applied magnetic field is increased. This behaviour  
  corresponds  to the entrance of  individual  quantized  vortices  
in the sample.  A quantitative understanding of the phenomena involved 
 can be achieved through  a numerical study of the Ginzburg-Landau equations 
  \cite{argentins,deo,deopeeters,palacios}.  
 Analytical progress can be made if one considers special limits  
 of these equations (such as the linear limit \cite{zwerger},  
 the London regime \cite{bb,vs} 
 or the Bogomol'nyi dual point \cite{am1}).  
 In later  experiments \cite{geim2}, it was  noticed 
 that when sweeping down the applied field 
 the sample exhibits a paramagnetic Meissner effect, 
 {\it i.e.} it has a paramagnetic magnetization. 
 Such a behaviour previously found 
 in high $T_c$ superconductors \cite{hightc} 
 had  been interpreted as a special features of these materials. 
  However, its occurrence in aluminum disks calls for  
a less specific explanation which results from  
the role played by metastable states at temperatures well below  
the critical temperature. 
 
 This work is devoted to  the study of  metastable 
states in a  mesoscopic superconductor.  
We shall investigate  vortex patterns in the 
London regime, {\it i.e.} in the limit $\kappa \gg 1$, where $\kappa$ 
is the   Ginzburg-Landau  parameter. Using  known   results \cite{bobel}
 for the magnetization of an infinite cylinder with a circular 
 cross-section  of radius $R$,  we derive,  in the mesoscopic 
 regime,  closed  expressions 
  for the magnetic field and the free energy  
 that are  suitable for analytical calculations.  
 They  allow   to obtain  the Bean-Livingston barrier and  to calculate 
 the  series of matching fields  (critical fields corresponding to  
  the entrance of the $N$-th vortex). 
 The paramagnetic Meissner state can then be 
 explained using  a simple  scenario for the hysteretic behaviour of 
 the magnetization, involving  the appearance and the disappearance  
 of the surface barrier.  
 
Once  the existence of metastable states 
 established, a criterion to classify them is required. The number 
 of vortices is a simple topological number but it is not precise  
  enough to  distinguish  between  all  different type of states. Projection 
 methods into eigenstates of  the linearized theory  
 produce  a set of integers   that were  proposed as classifying  
 numbers to  characterize a given  metastable vortex state \cite{palacios} 
 and  that  would generalize quantum numbers to a non-linear theory.  
 However, these numbers are non-generic and  
 it is not obvious to relate them   to the  geometric 
 features of a given  pattern. 
 A  more  geometrical 
 path will be followed here  to present  
 a topological description of  vortex patterns. 
 Our study is   based on an analogy 
 between  a configuration of vortices and  a dynamical system, obtained  by  
   interpreting the  superconducting current  
  as the local (phase-space) velocity of a  particle. The Hamiltonian of the associated 
 dynamical system   turns out to be the magnetic field. 
 We  shall  show that a configuration of vortices  
 can be characterized  by the critical points of  
 the magnetic field (or equivalently  the points where the current vanishes). 
 The number  of critical points of a given type 
 (maxima, minima, saddle-points)  will provide a natural  set of  topological 
 invariant integers  associated with a configuration.  
 This   geometrical analysis  reveals  the existence of structural   
 transitions  between states with the {\it same} number of vortices:  
 indeed,  the number and the nature of the 
 critical points  can vary  abruptly when the applied field 
 exceeds some specific values, provoking topological phase transitions. 
 For a simple polygonal configuration of vortices, we calculate 
 the  corresponding transition fields.  
 These topological changes are  in fact  best  visualized 
  using    a    `dual' description  
 of  a system of vortices, 
  closely  related to the paramagnetic 
 Meissner effect:  there exists  a   special closed curve  $\Gamma$,  
 first introduced in  \cite{am1}, 
 which is everywhere orthogonal to the current lines and which   
   passes  through  the critical points 
of the system. Therefore any structural  change affecting  the critical 
 points, will 
 manifest  itself in a topological change of  the shape 
 of  $\Gamma$. This  curve  
 thus  provides   an efficient and elegant tool to characterize 
  vortices  patterns and to understand 
 qualitatively the topological transitions  between different configurations. 
 
 The plan of this article goes as follows. In section 2, we recall how 
 the London free energy  can be obtained as a limit  of the  
 Ginzburg-Landau  theory. In section 3,  
 we present the solution of 
 London equation in the case of an infinite circular  cylinder  
(details of the calculations are given in Appendix A and B). 
 Section 4 is devoted  to the mesoscopic  limit:   we calculate  
   the magnetic field and the free energy,   study 
  the Bean-Livingston barrier and  obtain 
 the matching fields. The paramagnetic Meissner state is obtained 
 in section 5. Section 6 
 is devoted to the  topological investigation of vortex patterns: 
  we  show that the critical points  characterize 
 a configuration and  we demonstrate  the existence of  
  topological phase transitions. 
 In the second part of section 6, the dual description 
  of a configuration of vortices, via the  curve  
 $\Gamma,$  is introduced    to  provide 
  a better understanding of the transitions between patterns. 
 Section 7 contains some concluding remarks and discussions.

\section{The London limit of the Ginzburg-Landau equations}

Consider an infinite superconducting cylinder with a cross-section  
of area $\Omega $  
in the constant magnetic field $h_e \hat{z}$ along the axis of the cylinder. 
In the presence of a magnetic field  the Ginzburg-Landau free energy density 
per unit length of the cylinder,  defined as difference of free energies  
with and without magnetic field ${\cal F}=F_S(h)-F_S(0)$,  
reads  
\begin{equation} 
{\cal F}  =  
\int \left( {\vec{h}^2 \over 2} + \kappa^2  
 \left(1 - |\psi{|^2}\right)^2 +  
\left|\left({\vec{\nabla}} - i{\vec A}\right)\psi \right|^2\right)d^2 r 
\label{freeen} 
\end{equation} 
where $\psi = |\psi| {e^{i \chi}}$ is the order parameter and 
${\vec h} = {\vec \nabla} \times \vec A $ is  the local magnetic induction.  
The  
integration is performed over the cross-section of the cylinder. 
A superconductor is characterized by two length scales: the London penetration depth $\lambda$ 
and the coherence length $\xi$. The Ginzburg-Landau parameter is defined as their ratio  
$\kappa=\lambda/\xi$. 
 In this work lengths are measured in units of $\lambda \sqrt 2$ and 
the magnetic induction  in units of  the thermodynamic critical field 
$H_c =\phi_0/2\sqrt{2}\pi\lambda\xi $, where $\phi_0=hc/2e$ is the magnetic flux quantum. 
The free energy (\ref{freeen}) is given in units of $\xi^2 H_c^2/4\pi$.

The Ginzburg-Landau equations for $\psi$ and $\vec h$ are obtained from a  
variation of $\cal F$. They are given by 
\begin{eqnarray}  
  -  \left({\vec{\nabla}} - i{\vec A}\right)^2 \psi & = &  2{\kappa}^2 \psi  
\left(1 - |\psi{|^2}\right)   \label{adgl1} \\  
{\vec \nabla} \times {\vec h} & = & 2 |\psi |^2 \left({\vec \nabla } \chi - {\vec A}\right) 
  \label{adgl2}  
\end{eqnarray}  
 Equation (\ref{adgl2})  is the   Maxwell-Amp\`ere  
 equation which could be written as well using the current density  
${\vec \jmath}= \mbox{Im} \left( {{\psi^*}} {\vec \nabla} \psi \right) -  
|\psi{|^2}{\vec A}$.   
    Outside the superconducting sample,  
  $\psi = 0.$  The boundary condition  
 on the surface of the superconductor   is obtained by requiring  
 that the normal component of the current density vanishes  
 (superconductor/insulator boundary condition  \cite{degennes}):  
\begin{equation}  
\left.\left(\vec{\nabla} - i \vec{A}\right)\psi \right|_{\hat{n}} = 0  
\label{bondcond}  
\end{equation}  
 here ${\hat{ n}}$ is the unit vector  
 normal at each point to  the surface  
 of the superconductor.  
 
We study a superconductor in an applied external field $h_e$  
the relevant thermodynamic potential  is the Gibbs free  
energy $G$ obtained from $F$ via a Legendre transformation  
\begin{equation} 
G = F - \Omega h_e B  
\label{legendre} 
\end{equation} 
where  the total (dimensionless) magnetic induction (in the $\hat{z}$ direction )  
 $B$ is given by the averging 
the magnetic induction over the cross-section  
\begin{equation} 
 B = {1 \over \Omega} \int  h (\vec r) \,d^2 r 
\end{equation} 
where the magnetic induction is always in $\hat{z}$ direction: $\vec{h}=h\hat{z}$.

In the normal phase, we have $\psi = 0$, $B = h_e$ and $F_S(h)=F_N(h)=\Omega h^2/2$.  
Therefore, the corresponding Gibbs free energy $ G_N$  is  given by:  
\begin{equation}  
G_N  =  F_N (h)- \Omega h_e B =F_N(0) - \frac{\Omega h_e^2}{2} 
\label{gibbsnormal}  
\end{equation}  
 
At  thermodynamic equilibrium, the superconductor  selects the  state of  
minimal  Gibbs free energy ${\cal G}$. The  quantity which is measured in  
experiments is the magnetization   
$M$ of the superconductor due to  the applied field given by $4 \pi M =   
B - h_e $.  It is obtained, at thermodynamic equilibrium using  the  
thermodynamic relation \cite{degennes}:  
\begin{equation}  
 -  M = \frac{1}{2\pi}  
 \frac{\partial {\cal G}}{\partial h_e}  
 \label{magthermo}  
\end{equation}  
where the difference of the (dimensionless) Gibbs energies ${\cal G} =   G_S - G_N$ 
up to a constant equal to the superconducting condensation energy, is given by 
\begin{equation}  
{\cal G} = {\cal F} - \Omega h_e B  + \Omega  h_e^2/2 
\label{gsgn}  
\end{equation}

The dimensionless ratio $\kappa=\lambda/\xi$ is the only free parameter  
to describe the superconducting state. It determines, in the limit of an infinite  
system, 
 whether the sample is a type-I or type-II superconductor \cite{degennes}. 
For $\kappa \geq {1 / \sqrt 2}$, i.e. for type-II superconductors, it is energetically  
favorable for the  
system in a large enough magnetic field to sustain normal regions in the bulk which  
appear as vortices. 
Their magnetic flux quantized in units of $\phi_0$. We shall now study the system of  
vortices in an extreme 
Type II superconductor, {\it i.e.} in the London limit $\kappa\gg 1$.  
 
In this limit, the amplitude of the order parameter  
$|\psi|$ does not vary and can be taken to one except at the position of  
the vortices where it vanishes. The Ginzburg-Landau free energy reduces to  
\begin{equation} 
{\cal F}  = 
\int\left(\frac{ h^2}{2}  +  
\left|\left(\vec{\nabla} - i\vec {A}\right)\psi \right|^2\right) d^2 r 
\end{equation} 
Using this equation and   $|\psi|^2=1$, except at the position of the vortices, we  
can write 
\begin{equation} 
\left|\left(\vec{\nabla} - i\vec{A}\right)\psi\right|^2 =  
\left(\vec{\nabla} \times \vec {h}\right)^2 
\end{equation} 
so that the free energy is given by the expression 
\begin{equation} 
{\cal F}  =  
\frac{1}{2}\int \left(h^2  +\frac{1}{2}  
\left(\vec{\nabla} \times \vec{ h}\right)^2\right) d^2r 
\label{freeenergy} 
\end{equation} 
which in our units is independent of $\kappa$.  
To obtain the equation for $h$, we notice that the phase  
$\chi$ is non singular except near a vortex (say at $r=0$) where  
it has the property that 
\begin{equation} 
\oint {\vec \nabla} \chi\cdot {\vec dl} = 2 \pi 
\end{equation} 
on any closed contour encircling the singularity at point $r=0$. Using Stokes  
theorem,  
\begin{equation} 
\oint {\vec \nabla} \chi\cdot {\vec dl} =  
\int\!\!\!\int d{\vec S}\cdot {\vec \nabla} \times {\vec \nabla} \chi = 2 \pi, 
\label{stokes} 
\end{equation} 
we deduce that the $\hat z$-component of  
$\vec{\nabla} \times \vec{\nabla}\chi$ equals  $ 2 \pi \delta (\vec r  )$. 
 The equation for $h$ is then given by \cite{rem} 
\begin{equation} 
\nabla^2 h - 2h = - 4\pi \delta (\vec{r})  
\label{eqh1} 
\end{equation} 
This equation can be generalized to the case of $N$ vortices placed at  
points $\vec r_k$  
\begin{equation} 
\nabla^2 h -2 h = -4\pi\sum_{k=1}^{N} \delta ({\vec r} - {\vec r_k} ) 
\label{eqh2} 
\end{equation} 
In the next section we shall present the solution of this equation in the particular  
case of a cylinder  
with circular cross section.  
 
\section{ Solution for $N$ vortices in  a circular cylinder}   
 
>From now on we shall study a circular cylinder of radius $R$.  
The magnetic induction and Gibbs energy of a system of $N$ vortices in  
this geometry has been studied first in \cite{bobel}.  
The detailed derivation of the results  is presented in the Appendix A. 
The solution $h(r, \theta)$  
of the equation (\ref{eqh2}) satisfying the boundary condition $h(R,\theta)=h_e$ can be written as a  
sum of three terms: 
\begin{equation} 
h(r,\theta)=h_M (r) + h_V (r,\theta) + h_{\bar{V}}  (r,\theta) 
\label{3term} 
\end{equation} 
The first term  
\begin{equation} 
h_M (r) =h_e I_0 (\sqrt{2}r)/I_0 (\sqrt{2}R), 
\label{hM} 
\end{equation} 
where $I_0 (x)$ is the modified Bessel  
function of first kind, describes the Meissner effect in  
the absence of vortices. The second term is the magnetic induction of  
$N$ vortices placed at points $\vec{r}_k=(r_k,\theta_k)$ for $k=1,\ldots,N$ and it  
is given by 
\begin{equation} 
h_V (r,\theta)=2\sum_{k=1}^N K_0 (\sqrt{2}|\vec{r}-\vec{r}_k|) 
\label{hV} 
\end{equation} 
The third term, written with help of the modified Bessel function of first and second kind  
$I_n (x)$, $K_n (x)$ as 
\begin{equation} 
h_{\bar{V}} (r,\theta) = -2 \sum_{n=-\infty}^{+\infty} 
K_n (\sqrt{2}R)\frac{I_n (\sqrt{2}r)}{I_n (\sqrt{2}R)}  
 \sum_{k = 1}^{N}I_n (\sqrt{2}r_k) \cos n ( \theta - \theta_k)  
\label{hVbar} 
\end{equation} 
ensures the boundary condition  $h(R,\theta)=h_e$. We are interested in polygonal rings of  
vortices. For this particular configuration $r_k=d$,  
$\theta_k = 2\pi k/N$ and the expression for  $h_{\bar{V}} (r,\theta)$ can be simplified 
\begin{equation} 
h_{\bar{V}} (r,\theta) = -2 N \sum_{n=-\infty}^{+\infty} 
K_n (\sqrt{2}R) \frac{I_n (\sqrt{2}r) I_n (\sqrt{2}d) }{I_n (\sqrt{2}R)}  
\cos Nn \theta  
\label{hVbarp} 
\end{equation} 
 
The corresponding  
Gibbs energy has also been obtained in \cite{bobel}. The expression  
of the dimensionless Gibbs energy is given by 
\begin{equation} 
{\cal G} = {\cal F} -  h_e \int h({\vec r}) \,d^2 r+  \pi R^2 h_e^2/2 
\label{gibbs} 
\end{equation} 
The field $h_V$ diverges at the positions of the vortices and the corresponding part  
of the free 
energy must be regularized. The singular contribution coincides with one calculated   
by Abrikosov \cite{abrikosov} for  the case of infinite system and is given by   
\begin{equation} 
{\cal F}_\infty =N{\cal E} +2\pi\sum_{j\neq k}K_0(\sqrt{2}|\vec{r}_j - \vec{r}_k|) 
\label{f1} 
\end{equation} 
where ${\cal E}=2\pi(\ln\kappa+0.081)$ is the one-vortex energy. 
 
The regular part of the Gibbs energy  is calculated in the Appendix \ref{apgibbs}.   
Using the expressions (\ref{hM}),(\ref{hV}) and (\ref{hVbar}) for the magnetic field,  
one obtains  
\begin{eqnarray} 
{\cal G}_N  - {\cal G}_0 & = &N{\cal E} +2\pi\sum_{j\neq k}K_0(\sqrt{2}|\vec{r}_j - \vec{r}_k|) - 2 \pi h_e  
\sum_{k = 1}^{N} \left( 1 - { I_0 (\sqrt{2}r_k) \over I_0 (\sqrt{2}R)} \right) \nonumber \\  
& - & 2 \pi  \sum_{n=-\infty}^{+\infty}   \sum_{i,j=1}^N  
  \frac{K_n(\sqrt{2}R) I_n(\sqrt{2}r_i)I_n((\sqrt{2}r_j) }{ I_n(\sqrt{2}R)}  \cos{n}(\theta_{i} - \theta_j)  
\label{gibbse} 
\end{eqnarray} 
where ${\cal G}_0  = \pi R h_e^2\left(\frac{R}{2}-\frac{\sqrt{2}}{2} 
\frac{I_1(\sqrt{2}R)}{I_0(\sqrt{2}R)}\right)$  
is the Gibbs energy in the absence of vortices {\it i.e.} in the Meissner state. 
For the particular case of vortices distributed on a regular polygon of radius $d$,  
the previous expression (\ref{gibbse}) simplifies and becomes  
\begin{eqnarray} 
 {{\cal G}_N -  {\cal G}_0 \over N} & = & 
 {\cal E} + 2 \pi  \sum_{l=1}^{{N-1}}  K_0(|2\sqrt{2}d\sin\frac{l\pi}{N}|)   
   - 2 \pi  h_e    \left( 1 -  \frac{I_0(\sqrt{2}d)}{ I_0(\sqrt{2}R)}  \right)\nonumber\\  
 & - &  2 \pi  N  \sum_{n=-\infty}^{+\infty}  
 \frac{K_{Nn}(\sqrt{2}R) I_{Nn}(\sqrt{2}d)^2}{ I_{Nn}(\sqrt{2}R)} 
\label{gibbspolygon} 
\end{eqnarray} 
Similar expressions have been already obtained previously in \cite{bobel,vs} and we  
shall now discuss them in the limit  
of a small radius $R$ in order to use the particularly elegant structure of the  
corresponding solutions  
to derive analytical expressions for the field and the energy.

\section{The limit of a small disk} 
 
We shall now consider the limiting case  
where $R \ll \lambda $ for which the corresponding Gibbs free energy was obtained 
in \cite{bb}. In this limit 
 the previous expressions greatly simplify. Since all the distances are given  
in units of $\lambda\sqrt{2}$, we can expand  
the expression for the magnetic induction using the  expansion of the Bessel functions  
for small argument, so that 
\begin{equation} 
 h (z,\bar{z}) - h_e  = \phi_e(|z|^2 - 1) + 2\sum_{i=1}^N \ln \left|\frac{ 1 - \bar{z_i} z}{z - z_i}\right| 
\label{hlim} 
 \end{equation}  
where $\phi_e = h_e R^2/2$ is the flux of the external field through the system and  
we use the complex notations, 
\begin{equation} 
  z = \left(\frac{r}{ R}\right) e^{i\theta}  \,\,  \hbox{ and   } \,\, 
  z_k = \left(\frac{r_k}{ R}\right) e^{i\theta_k} 
\end{equation} 
together with  complex conjugate  $\bar{z}$ and $\bar{z}_k$.  
This expression for the magnetic induction is solution of the modified London equation 
\begin{equation} 
{\partial^2 \over \partial z \partial \bar{z}} h (z,\bar{z}) -  h_e = - \pi \sum_{i=1}^{N}  
\delta (z - z_i )\delta (\bar{z} - \bar{z}_i ) 
\label{eqhz} 
\end{equation} 
which can be obtained  from (\ref{eqh2})  for small  variation of the field on the scale  
of $R$.

In the particular case, we shall consider from now on, of vortices 
 distributed along a regular polygon of radius $d$, we have $r_k =d $ and for the $N$ 
 vortices sitting on a shell,  
the angle $\theta_k$ of the $k$-th vortex is $\theta_k = 2 \pi k/ N$.  
Then, defining the real quantity $x = d/R \leq 1$, the relation (\ref{hlim}) becomes  
\begin{equation} 
h_N (z,\bar{z}) - h_e = \phi_e (|z|^2 - 1) + 2\ln \left|\frac{1 - x^N z^N}{x^N - z^N}\right|  
\label{hdez} 
\end{equation} 
whereas the Gibbs energy (\ref{gibbspolygon}) becomes 
\begin{equation} 
{{\cal G}_N - {\cal G}_0 \over N}  =   
{\cal E}' - 2\pi\phi_e ( 1 - x^2) + 2\pi \ln(1 - x^{2N}) - 2\pi (N-1) \ln x - 2\pi\ln N 
\label{gibbs2} 
\end{equation} 
where ${\cal G}_0 = \pi \phi_e^2/2$ and ${\cal E}'\simeq 2\pi \ln \left(R/\xi\right)$. 
 
The relation between the radius $x$ of the polygonal configuration and the  
external flux $\phi_e$  
is obtained by minimizing ${\cal G}_N (x, h_e)$ with respect to $x$ at fixed $\phi_e$, namely 
\begin{equation} 
\phi_e =  \frac{N-1}{2 x^2} + \frac{ N x^{2N-2}}{1 - x^{2N}}  
\label{xequilib} 
\end{equation} 
 
\subsection{The Bean-Livingston barrier  \label{bl}} 
 
\begin{figure} 
{\hspace*{-0.2cm}\psfig{figure=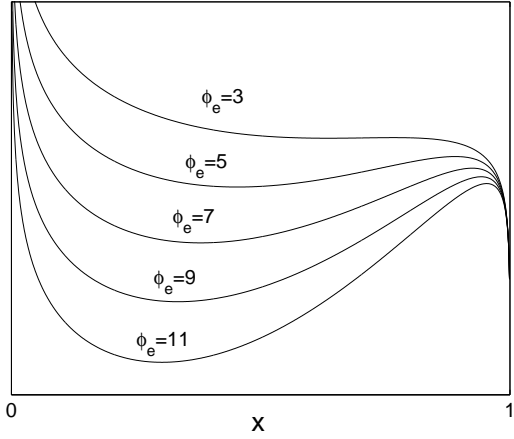,height=6cm,angle=0}} 
{\vspace*{.13in}} 
\caption{Behavior of the Gibbs energy as a function of the position $x$  
of the $N=3$ vortex shell. The maximum which exists for high enough  
applied field corresponds to the unstable equilibrium point  
for the vortex configuration and gives the height of the Bean-Livingston barrier.} 
\label{Bl} 
 
\end{figure}

The relation (\ref{xequilib}) has, as a function of the applied flux $\phi_e$,  
either zero, one or two solutions. The latter  
case corresponds for the Gibbs energy (\ref{gibbs2}) to the existence of a Bean-Livingston  
potential barrier \cite{bl} for which one solution is  
stable and gives the equilibrium position of the vortices while the second  
solution is unstable and gives the height  
of the potential barrier as shown in the Fig. \ref{Bl}. It should be noticed that for  
more than one vortex $(N > 1)$, the energy ${\cal G}_N$ diverges logarithmically at the 
 origin $x =0$ due to the repulsion between the vortices.  
 
\begin{figure} 
{\hspace*{-0.2cm}\psfig{figure=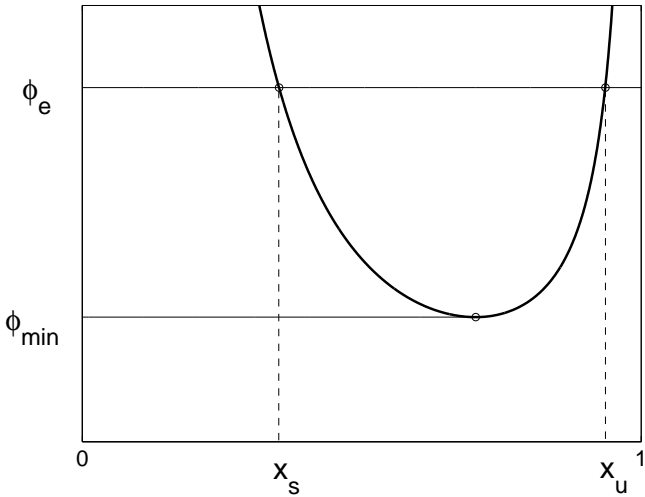,width=8.6cm,angle=0}} 
{\vspace*{.13in}} 
\caption{Graphical solution of equation~(\ref{xequilib}). The curve shows 
the behavior of $\phi_e (x)$. The two solutions $x_s$ and $x_u$ correspond  
respectively to the stable and unstable positions of the vortex shell for  
values of the magnetic field above $\phi_{min}$. } 
\label{Hmin} 
 
\end{figure}

The potential barrier disappears  
for low enough applied flux below a characteristic value $\phi_{min}$ given by the minimum  
of the function $f_N (x)$ in the rhs of (\ref{xequilib}) as  
represented in Fig. \ref{Hmin}. Defining $y = x^2$, we have  
\begin{eqnarray} 
f_N(y)&=&\frac{N-1}{2y}+\frac{Ny^{N-1}}{1-y^N}\\ 
f'_N (y) &=& -\frac{N-1}{2y^2} + \frac{N(N-1)y^{N-2}}{ 1 - y^N}+\frac{N^2 y^{2N-2}}{(1-y^N)^2} 
\end{eqnarray} 
For large values of $N$, we have $y_{min}^N \simeq 1/ 2 N$ so that  
\begin{equation} 
\phi_{min} \simeq \frac{N-1}{2\sqrt[N]{2N}} \simeq \frac{ N-1}{2}  
\label{phimin} 
\end{equation} 
For $\phi_e < \phi_{min}$, there is no stable solution and then no equilibrium  
position for $N$ vortices exists. 
 
\subsection{Matching fields} 
 
\begin{figure} 
{\hspace*{-0.2cm}\psfig{figure=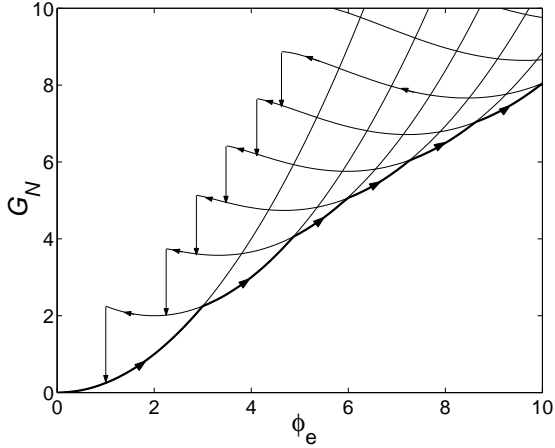,height=6cm,angle=0}} 
{\vspace*{.13in}} 
\caption{Dependence of the Gibbs energy as a function of the applied flux  
$\phi_e$. The equilibrium behavior of the system is described by the envelop  
of the curves for ${\cal G}_N (h_e)$. Arrows show the direction in which the magnetic  
flux is varied. } 
\label{Gn} 
\end{figure}

By plugging the expression (\ref{xequilib})  
 into the expression (\ref{gibbs2}), we obtain the behavior of the Gibbs energy  
for $N$ vortices as a function of the applied flux $\phi_e$.  
This expression is the counterpart,  
in the London limit, of the parabolae obtained at the Bogomoln'yi point \cite{am1}  
for $\kappa = 1/ \sqrt 2$. For a given value $\phi_e$ of the applied  
field, the optimal number of vortices corresponds to the envelop of the set of curves  
${\cal G}_N (\phi_e)$ as represented in the Fig. \ref{Gn}. The matching values  $\phi_N$ are  
the values taken by the flux of 
 the external field at which ${\cal G}_{N-1} (\phi_N) ={\cal G}_{N} (\phi_N)$.  
They correspond to the transition  between configurations with $N-1$ and $N$ vortices  
respectively.  

\begin{figure} 
{\hspace*{-0.2cm}\psfig{figure=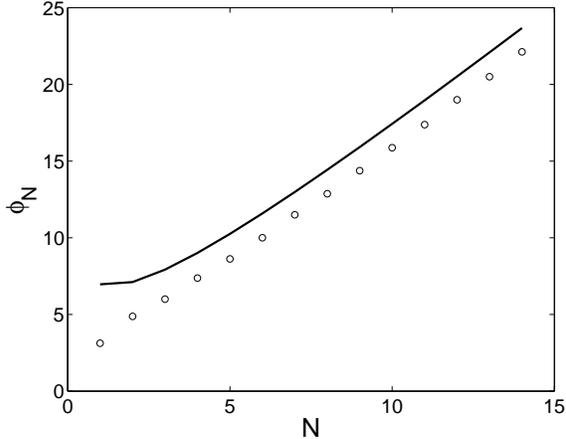,height=6cm,angle=0}} 
{\vspace*{.13in}} 
\caption{Comparison between the relation (\ref{matchingnum}) for the matching fields  
$\phi_N$ and the  numerical results for $\phi_1=\ln \left(R/\xi\right)=3$. } 
\label{hn} 
\end{figure}

Using the relation (\ref{gibbs2}), it is straightforward to obtain the  
field at which the first vortex penetrates the system; it is given by  
$\phi_1= {\cal E}'/2\pi = \ln\left(R/\xi\right)$. This   
field is larger than the field at which the Bean-Livingston potential barrier disappears. 
 For larger values of $N$, in the limit $\phi_e\gg \phi_{min}$, we approximate the  
relation (\ref{xequilib}) by  
$\phi_e = (N-1)/2 x^2$ so that  
\begin{equation} 
\frac{{\cal G}_N -{\cal G}_0}{2\pi} 
\approx   
N \left(\phi_1-\phi_e\right) +  
\frac{N (N-1)}{2}\left(1+\ln\frac{2\phi_e}{N-1}\right)- N\ln N 
\end{equation} 
Equating this to the similar expression for ${\cal G}_{N-1}(\phi_e)$ we obtain 
that the matching fields for $N\rightarrow\infty$ are given by solution  
of the following equation 
\begin{equation} 
\phi_N=\phi_1+\frac{N-1}{2}\left(1+2\ln\frac{2\phi_N}{N}\right)-\ln N+1+{\cal O}(1/N) 
\label{matching} 
\end{equation} 
We use the following ansatz $\phi_N=\frac{1}{2}(a N+b\ln N+c)$. Substituting 
it to (\ref{matching})  
we obtain  the  coefficients  
\begin{equation} 
b=-\frac{2a}{a-2},\;\;\;\;\;\;c=\frac{2a}{a-2}\left(1+\phi_1-\frac{a}{2}\right), 
\label{bandc} 
\end{equation} 
where the coefficient 
$a$ satisfies the transcendental equation: 
\begin{equation} 
a=1+2\ln a 
\end{equation} 
This equation has one obvious solution $a=1$ which is compatible with  
(\ref{matching}). However from the  exact numerical calculations 
of the matching fields using the relation (\ref{xequilib}) 
 we infer that the second root $a\simeq 3.5129$ is realized in the  
solution for the matching fields. It leads to the formula  
\begin{equation} 
\phi_N = 1.76 N-2.32 \ln N +2.32\phi_1-1.76 +{\cal O}((\ln N)^2/N)
\label{matchingnum} 
\end{equation} 
In the Fig.~\ref{hn} the values of the matching fields 
obtained from the computed free energy are compared with our prediction. 
The discrepancy is attributed to the slow convergence due to the term ${\cal O}((\ln N)^2/N)$. 

\subsection{Magnetization and paramagnetic Meissner effect} 
 
A paramagnetic behavior for the total magnetization results from the existence of a  
high enough Bean-Livingston barrier and the long life-time of the corresponding metastable 
states. We propose the following scenario for the hysteretic behavior  
of the magnetization  when the magnetic field is increased and then swept down.  
 
\begin{figure} 
{\hspace*{-0.2cm}\psfig{figure=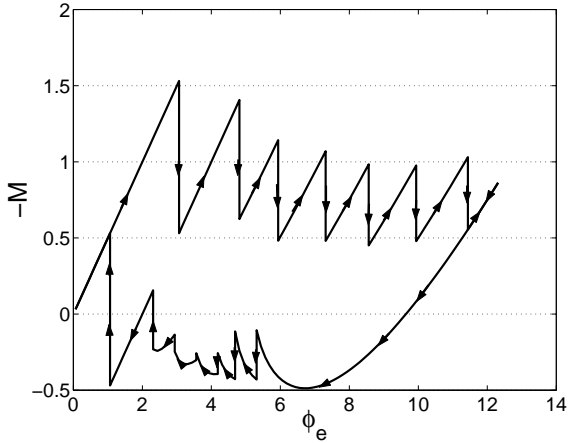,height=6cm,angle=0}} 
{\vspace*{.13in}} 
\caption{Magnetization of the disk as a function of flux $\phi_e$ for $\phi_1=\ln \left(R/\xi\right)=3$. 
 The hysteretic behavior is indicated by the arrows. } 
\label{magnet} 
\end{figure}

If thermal equilibrium is maintained while increasing slowly the external flux $\phi_e$, the  
results of the preceding section hold and the magnetization is given by the derivative of  
the envelop of the  
curves ${\cal G}_N (\phi_e)$ according to the thermodynamic relation (\ref{magthermo}).  
This behavior corresponds 
to the upper curve in the Fig.~\ref{magnet}.  
 
When the applied field is decreased, the vortices that are already in the system are 
confined by the Bean-Livingston barrier and cannot escape the  sample until the  
flux $\phi_e$ reaches a 
certain minimum value at which the barrier can be overcome, for instance, by the  
effect of  
thermal fluctuations.  
For the present discussion we use the ultimate criterion of zero barrier  as a  
definition of the  transition field.  
 
>From the discussion of section \ref{bl}, we deduced that the potential barrier  
for $N$ vortices  
disappears for a characteristic field given for large $N$ by the relation  
(\ref{phimin}),  
$\phi_{min} \sim (N- 1)/2$.  
The number of vortices is 
then decreased by one at this field and the energy drops as shown by the arrows in  
Fig.~\ref{Gn}. The magnetization calculated from (\ref{magthermo})  
 is shown by the lower curve in Fig.~\ref{magnet}. 
It  shows clearly a paramagnetic behavior, although it disagrees quantitatively with  
experiments and the numerical simulations based on the exact solution of the  
Ginzburg-Landau  
equations. This discrepancy results from the rather unrealistic London limit  we used for systems for  
which $\kappa\sim 0.3$ in order to obtain an  
analytic  expression for the Gibbs energy.  We emphasize that the system is not at  
thermal equilibrium but in  
a metastable state.   It is worth noticing that for  values of the parameter  
$\phi_1={\cal E}'/2\pi=\ln R/\xi\sim 2$,  
the paramagnetic Meissner effect can be observed as well for an equilibrium  
configuration, {\it i.e.} for a  minimum of the Gibbs  free energy. But this  
corresponds to $R\sim\xi$ so that the delta-like vortex approximation,  
crucial for the present discussion, is not  justified.

\section{Topological phase transitions between metastable vortex patterns} 
 
Up to now, we have considered vortex configurations for which the position  
$x$ of the vortex ring was given by (\ref{xequilib})
 resulting from the minimization of the  
energy. We shall now relax this condition and study 
the behavior of the metastable vortex  
patterns  as a function of the applied magnetic flux $\phi_e$. Physically,  
this corresponds to a situation where $N$ vortices are pinned by a   
 potential and remain on fixed positions.

 \subsection{An effective Hamiltonian system}

When varying the field $\phi_e$, transitions in a 
vortex pattern with  fixed number $N$   
of vortices will  appear. To characterize them, we shall study
 the behavior of the magnetic field $h$  
 given by (\ref{hdez}). To that purpose, it is interesting
 first to notice that, in two dimensions,  
the Maxwell-Amp\`ere equation  $\vec{\jmath}=\vec{\nabla}\times h \hat{z}$ 
is the Hamilton equation of  
a system whose generalized coordinates $(p,q)$ are
 the coordinates $(x,y)$ such that  
$j_x = \dot{x}$, $j_y=\dot{y}$ and the canonically conjugated 
momentum: 
\begin{equation} 
\left\{ \begin{array}{ccr} 
\dot{p} & = & -\partial_q h \\ 
\dot{q} & = & \partial_p h 
\end{array}\right. . 
\label{hameq} 
\end{equation} 
Then, the magnetic field corresponds to the
 Hamiltonian of the system and the current  
lines are the phase space trajectories. 
For a Hamiltonian system, the flow can be    characterized 
by its critical (fixed)   
points at which
\begin{equation} 
 {\vec \nabla} h = 0
\label{ptcritique} 
\end{equation}
Using (\ref{hameq}), we see that critical point correspond to
 zero velocity points 
such that $\vec{\jmath} =0$. 

 In such a description, a 
 vortex  corresponds to a maximum in  phase-space. However there also exist 
 in the system  other critical points, such as minima and saddle-points.  
 The number $N_M = N$  of maxima ({\it i.e.} vortices),
 $N_m$  of minima, and $N_s$ of saddles are not independent, because
 of the Euler-Poincare relation \cite{morse}, namely the 
 Euler-Poincar\'e characteristic  $\chi$ is given by
\begin{equation} 
\chi=N_M-N_s+N_m 
\label{EP} 
\end{equation}  
The integer  $\chi$ is a topological invariant and is
 equal to 1 for a disk thus providing a constraint between  the numbers  of 
the different types of  critical points.  
Hence, for a system with a given number of vortices,
 the difference $N_s-N_m$ between saddles and minima is fixed.
However,  each of them  
can vary and the set of these numbers provides a 
complete description of the topology of the vortex  
configurations, namely each topological phase can be described
using this set of integers.  
 
\begin{figure} 
{\hspace*{-0.2cm} 
\psfig{figure=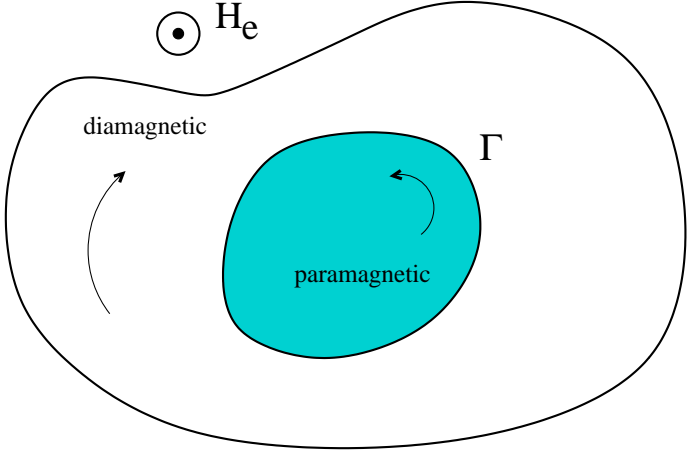,height=6cm,angle=0}} 
{\vspace*{.13in}} 
\caption{Paramagnetic and diamagnetic zones separated by the curve $\Gamma$.  } 
\label{paradia} 
\end{figure}

A different way to encode 
the position and the distribution of the critical
 points of the magnetic field in the disk is to use  
 a special  contour $\Gamma$,   introduced in \cite{am1}.  
 The curve  $\Gamma$    is defined by the condition  
that at each point $\vec r$ it is perpendicular
 to the current density $\vec{\jmath} (\vec{r})$. The equation  
of $\Gamma$ is then 
\begin{equation} 
{dr \over d \theta} = r^2 {\partial_r h \over \partial_{\theta} h} 
\label{eqgamma} 
\end{equation} 
To this definition we must add  
the requirement that $\Gamma$ must pass through the critical points at which  
$\vec{\jmath} (\vec{r})=0$. The curve
 $\Gamma$ has  several branches;  
one of them encircles all the vortices and defines a
 natural boundary between the diamagnetic (Meissner) domain 
near the boundary and the inner paramagnetic domain which includes the vortices (Fig.~\ref{paradia}).  
An example of this curve is shown in Fig.~\ref{gamma1} 
 in   the case of a disk with  one vortex.  
 
The study of the critical  
points and the resulting behavior of $\Gamma$
 provides a complete description of the vortex states either   
for thermodynamic or metastable states  
and of the transitions between different sets
 of topological numbers when varying the applied field.  
>From the behavior of the curve $\Gamma$ as a 
function of the applied field $\phi_e$,  
it might be interesting to draw  
the analogy with the pressure exerted on a
 closed membrane separating two  systems. 
 
 \subsection{Topological study of a vortex ring}

 We now    consider the simple case where
 all the vortices are placed on regular polygon at a distance $x$
 from the center of the disk. The
  critical points can be   classified (see Appendix C)
according  to their stability  
properties, obtained  from the hessian matrix  
$H_{ij}=\partial_i\partial_j h$.  We  review  the 
various topological structures  obtained and the transitions between
 them. 

 For a ring configuration of $N$ vortices
 located  at the distance $x$ from the center 
the general expression (\ref{hlim})  for the field 
$h$ in circular coordinates  
$r$, $\theta$ reads  
\begin{equation} 
h = \phi_e (r^2-1) +\ln \left(\frac{1-2 x^N r^N \cos N\theta+x^{2N} r^{2N}} 
{x^{2N}-2 x^N r^N \cos N\theta+ r^{2N}}\right) 
\label{hcirc} 
\end{equation} 
At the critical point the partial derivatives of $h$ vanish.  
The condition $\partial_r h = 0$ leads to  
\begin{equation} 
\phi_e r - 
N(x^{2N}-1)\, r^{N-1}\frac{x^N\cos N\theta\,(r^{2N}+1)-(x^{2N}+1)r^N} 
{\left(1-2x^Nr^N\cos N\theta+x^{2N}r^{2N}\right) 
\left(x^{2N}-2x^Nr^N\cos N\theta+r^{2N}\right)}= 0 
\label{dhdr} 
\end{equation} 
while  the condition 
$\partial_\theta h =0$  rewrites as
\begin{equation} 
N r^N\sin N\theta \frac{(1-x^{2N})(1-r^{2N})} 
{\left(1-2x^Nr^N\cos N\theta+x^{2N}r^{2N}\right) 
\left(x^{2N}-2x^Nr^N\cos N\theta+r^{2N}\right)}= 0 
\label{dhdtheta} 
\end{equation} 
 The solutions of equations (\ref{dhdr}) and 
  (\ref{dhdtheta}) are studied in detail in Appendix C.
One main feature is that one has to distinguish between  the cases
 $N=1$, $N=2$ and $N>2$ where $N$ is the number of vortices.
 We now describe these three cases separately.

\subsubsection{Metastable configurations for $N=1$}

For one single metastable vortex located at distance $x$ from the center  
there is no other critical point for weak flux $\phi_e<\phi_m(1)$, with   
\begin{equation} 
 \phi_m(1) =\frac{1-x}{1+x}. 
\label{pim1} 
\end{equation} 
At $\phi_e=\phi_m(1) $ a critical point appears at the point 
$r=1, \theta=-\pi$. For a flux slightly above $\phi_m(1) $,
 this point splits into a  
minimum at $r<1, \;\theta=-\pi$ and two saddle points on the boundary  
$r=1$ situated symmetrically about the diameter
 passing through  the vortex.  
As the flux increases the minimum moves towards
 the center and the saddle points  
approach the point $r=1,\;\theta=0$ each from its side. They coalesce at this  
point for  $\phi_e=\phi_M(1)$, where  
\begin{equation} 
\phi_M(1)   = \frac{1+x}{1-x}  
\label{piM1} 
\end{equation} 
and for larger values of the  
flux one saddle point appears inside the disk between the 
vortex and the boundary  
in the interval $x<r<1,\;\theta=0$. When $\phi_m(1)<\phi_e<\phi_M(1)$,
 the definition (\ref{EP})  
must be generalized to the case of critical point 
on the boundary \cite{morse}. It becomes  
\begin{equation} 
\chi=N_M-N_s+N_m-\frac{1}{2} N_{sb} 
\label{EPb} 
\end{equation}  
where $N_{sb}$ is the number of saddle points on the boundary. 
Thus the index is  again equal one. 
 
\begin{figure} 
{\hspace*{-0.2cm} 
a.\psfig{figure=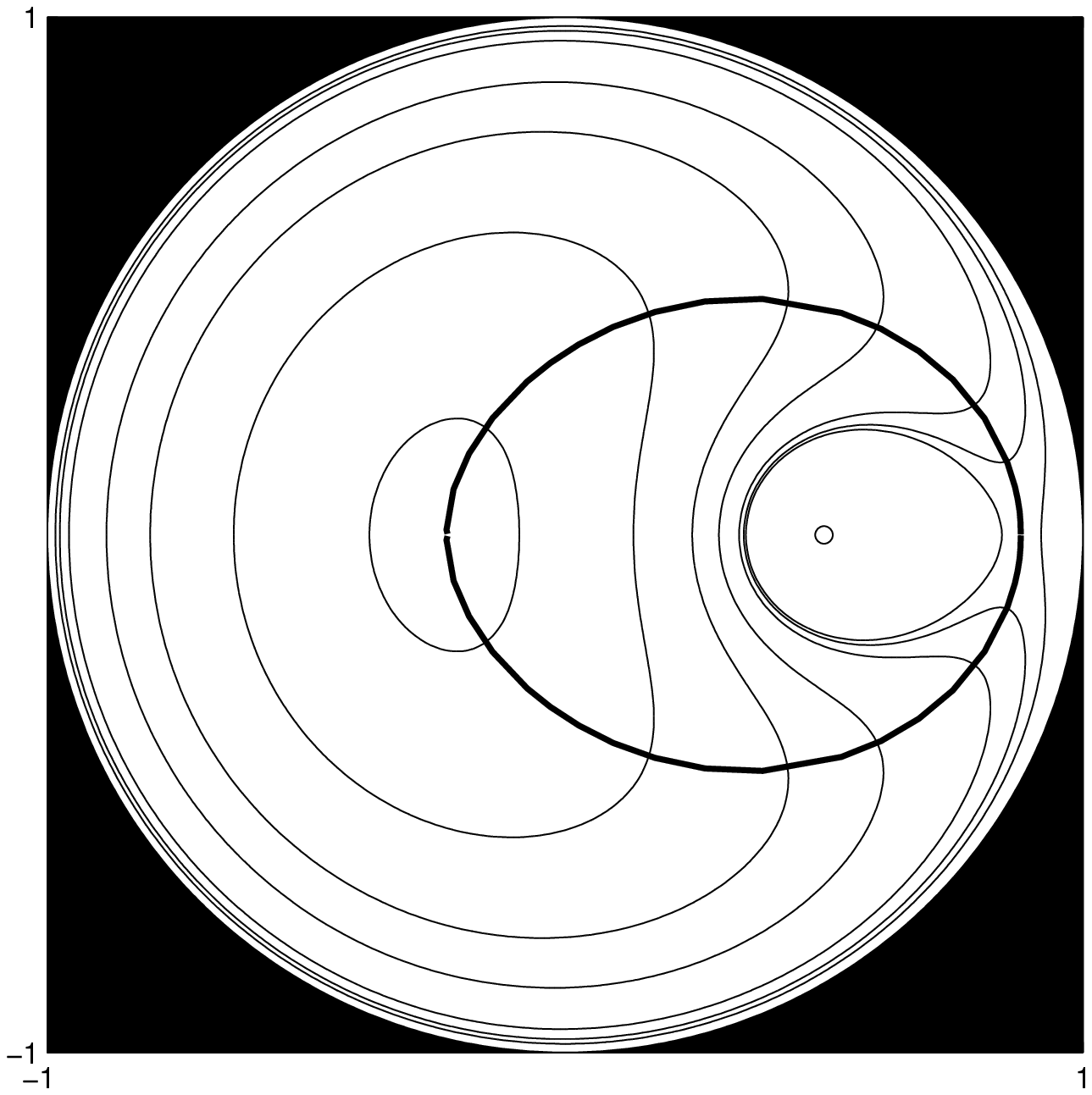,height=6cm,angle=0} 
\ \ \  b.\psfig{figure=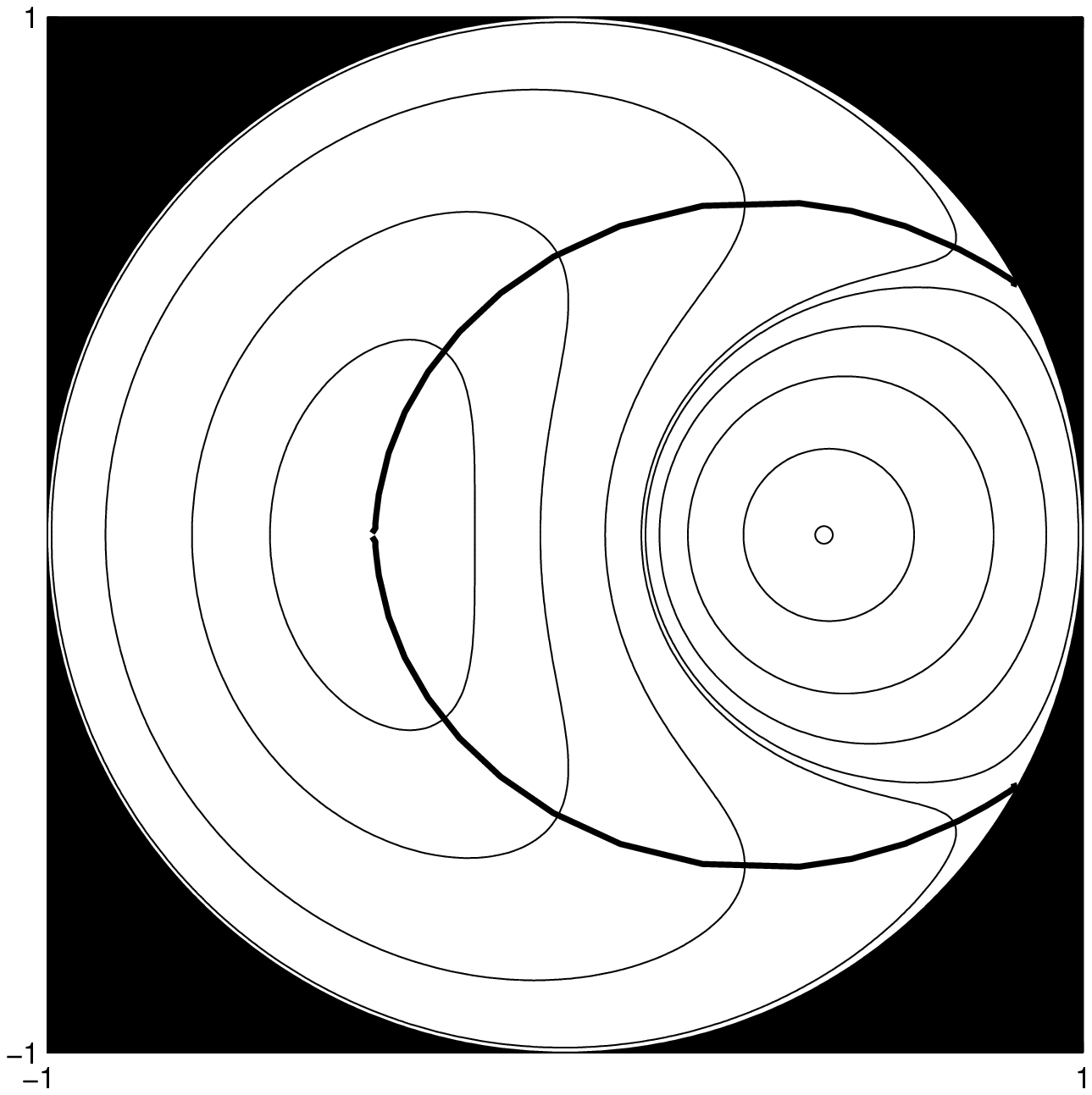,height=6cm,angle=0}} 
{\vspace*{.13in}} 
\caption{The case of one single vortex ($N=1$). 
Behavior of the curve $\Gamma$ for  a). $\phi_e > \phi_M(1)$
 and  b). $\phi_m (1)<\phi_e < \phi_M(1)$ } 
\label{gamma1} 
\end{figure}

An efficient  way to visualize these transitions is examine the changes in 
 the  curve $\Gamma$  as  the magnetic flux varies.  
In the case $\phi_e>\phi_M(1),$ shown in Fig.~\ref{gamma1}a, the curve
  $\Gamma$ 
lies inside the disk and separates the paramagnetic 
(internal) and diamagnetic (external) 
regions. For $\phi_e= \phi_M(1)$ the saddle point reaches 
 the boundary of the disk and for  
$\phi_m(1)<\phi_e<\phi_M(1)$ the curve $\Gamma$ starts and ends
 on the saddle points on the  
boundary (see Fig.~\ref{gamma1}b). 
Therefore, in this regime, one cannot define a  
paramagnetic region inside the disk  
and there is no closed curve that encloses a unit of flux.
 There appears an arc of circle   on  
the boundary where the direction of the current flow is opposite 
to the screening current 
responsible for the Meissner effect. At $\phi_e=\phi_m(1)$ all
 the disk becomes  
paramagnetic and the curve $\Gamma$ disappears.

\subsubsection{Metastable configurations for $N=2$} 
 
\begin{figure} 
{\hspace*{-0.2cm} 
a.\psfig{figure=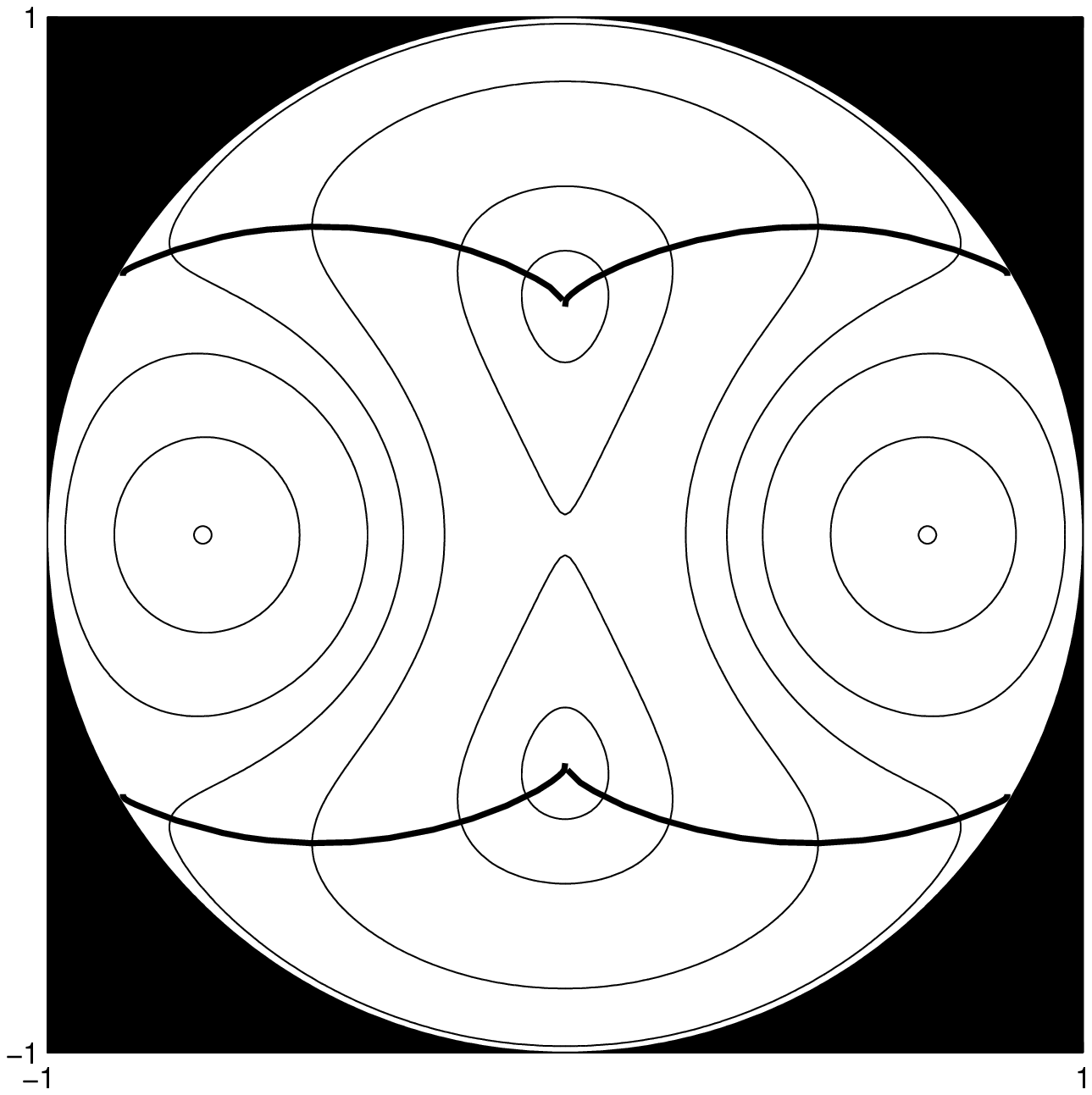,height=4.5cm,angle=0} 
\ \ \  b.\psfig{figure=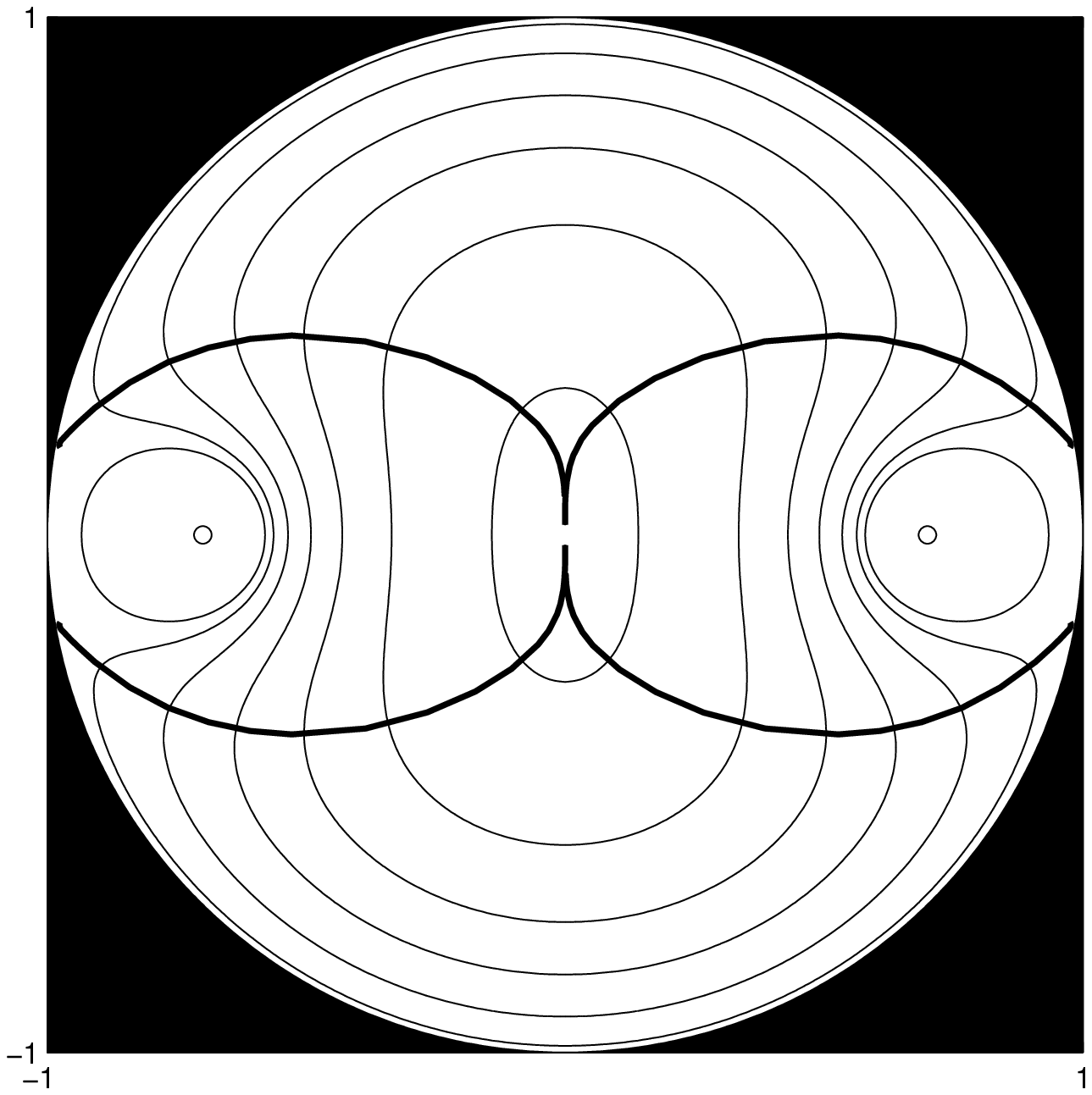,height=4.5cm,angle=0}
\ \ \ c.\psfig{figure=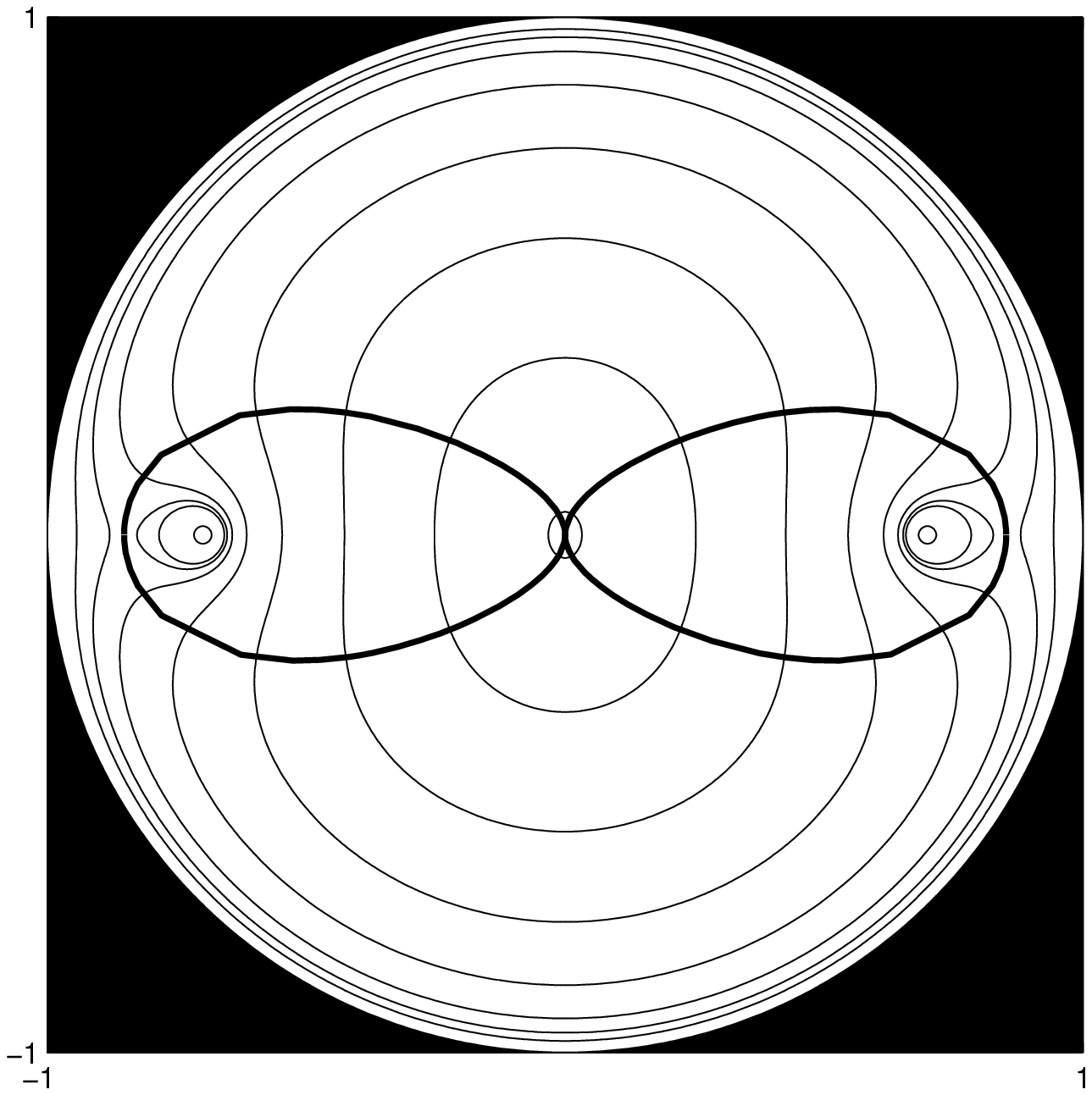,height=4.5cm,angle=0} } 
{\vspace*{.13in}} 
\caption{The case of $N=2$  vortices  for  a). $\phi_m (2)<\phi_e < \phi_c(2)$,
b).  $\phi_c (2)<\phi_e< \phi_M (2)$ and c).  $\phi_e > \phi_M(2)$.
The curve $\Gamma$ is shown. The position of the vortices is taken to be
$x=0.7>\sqrt{2-\sqrt{3}}$, therefore $\phi_c (2)<\phi_M (2)$.  } 
\label{gamma2} 
\end{figure}

For low enough field $\phi_e<\phi_m (2)$, with 
\begin{equation} 
 \phi_m(2) = 2\frac{1-x^2}{1+x^2}. 
\end{equation}
 the only critical point is a saddle point at the center.  
For $\phi_m(2)<\phi_e<\phi_M (2)$, where
\begin{equation} 
 \phi_M(2) = 2\frac{1+x^2}{1-x^2}. 
\end{equation}
 there are  two minima situated symmetrically on the diameter
 perpendicular to the line passing  
through the vortices and two pairs of saddle points
 on the boundary. This configuration is shown in~Fig.~\ref{gamma2}a.
Each pair is  
symmetric with respect to the line passing  through the  vortices and moves  
towards the points $r=1,\; \theta=0$ or $\theta=\pi$.
 For $\phi_e>\phi_M(2)$ each pair disappears  
and gives  birth to one single saddle point located
 between a vortex and the boundary  
at $\theta=0$ or $\pi$ as shown in~Fig.~\ref{gamma2}c.  
 
Two minima inside the disk move towards the center and at the critical 
value of the  
flux $\phi_e=\phi_c (2)$, where $\phi_c (2)=(1-x^4)/x^2$ they merge with the  
central point, which becomes a minimum for  $\phi_e>\phi_c(2)$. This critical  
value can be shown to be  always greater than $\phi_m(2)$, but it can be  
greater or less than $\phi_M(2)$ depending on the value of $x$.  
For $x<\sqrt{2-\sqrt{3}}\approx 0.52$,  one has $\phi_c (2)> \phi_M(2)$, 
while for $x>\sqrt{2-\sqrt{3}},$ we have $\phi_c (2) < \phi_M(2)$. The latter
case is shown in Fig.~\ref{gamma2}b and \ref{gamma2}c.

\subsubsection{The case $N>2$: topological phase transitions} 
 
\begin{figure} 
{\hspace*{-0.2cm} 
a.\psfig{figure=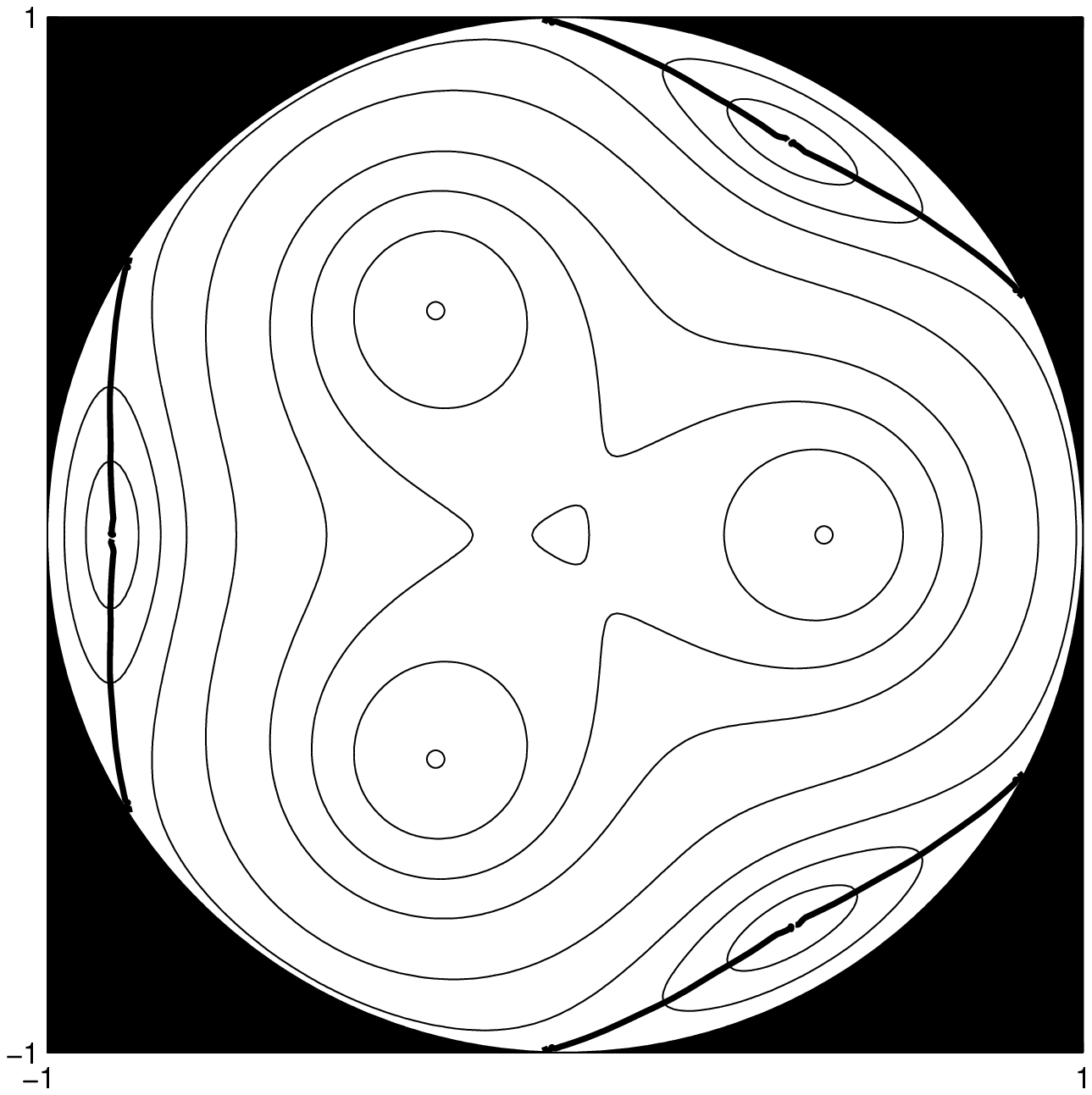,height=4.5cm,angle=0} 
\ \ \  b.\psfig{figure=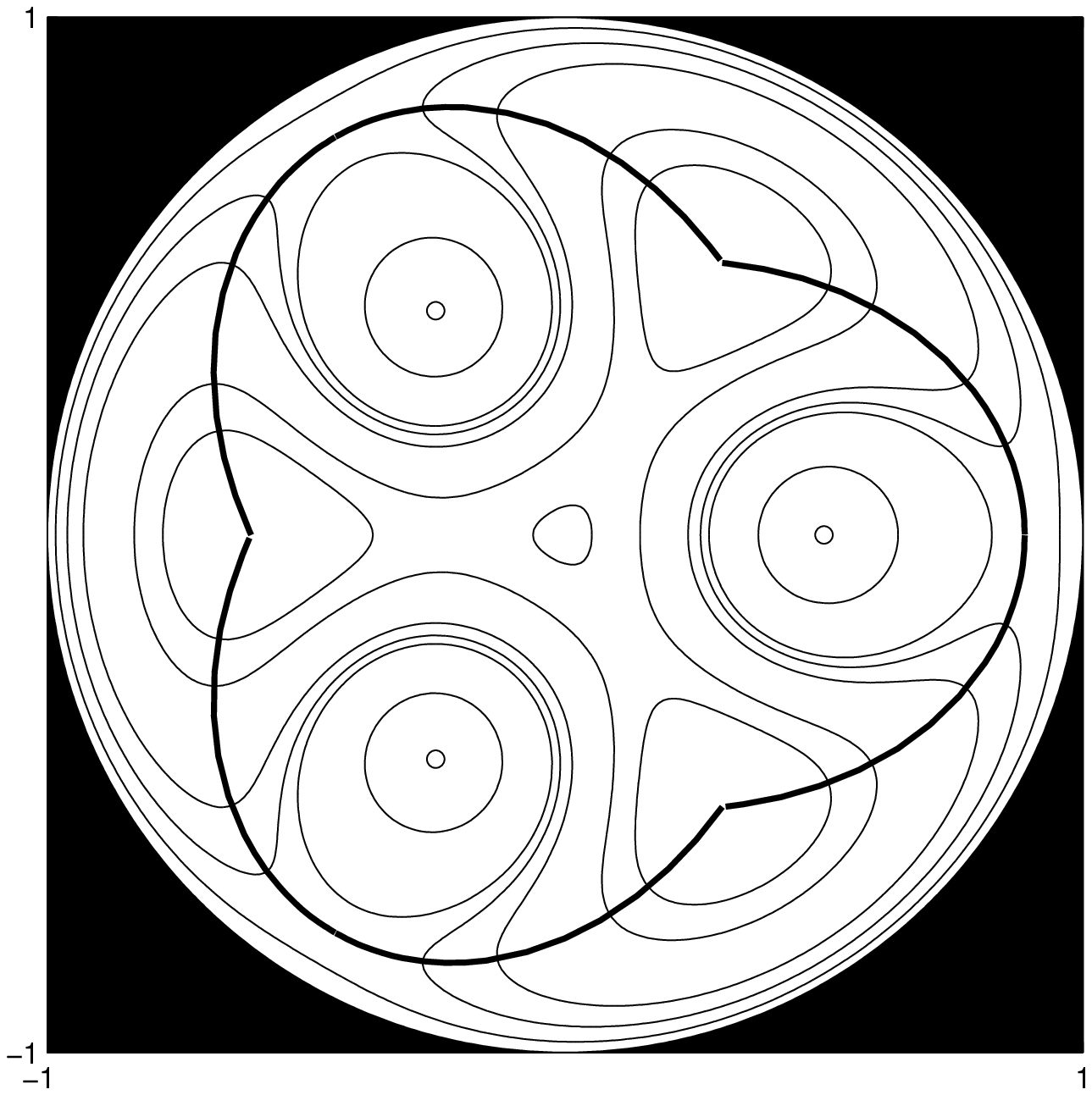,height=4.5cm,angle=0}
\ \ \ c.\psfig{figure=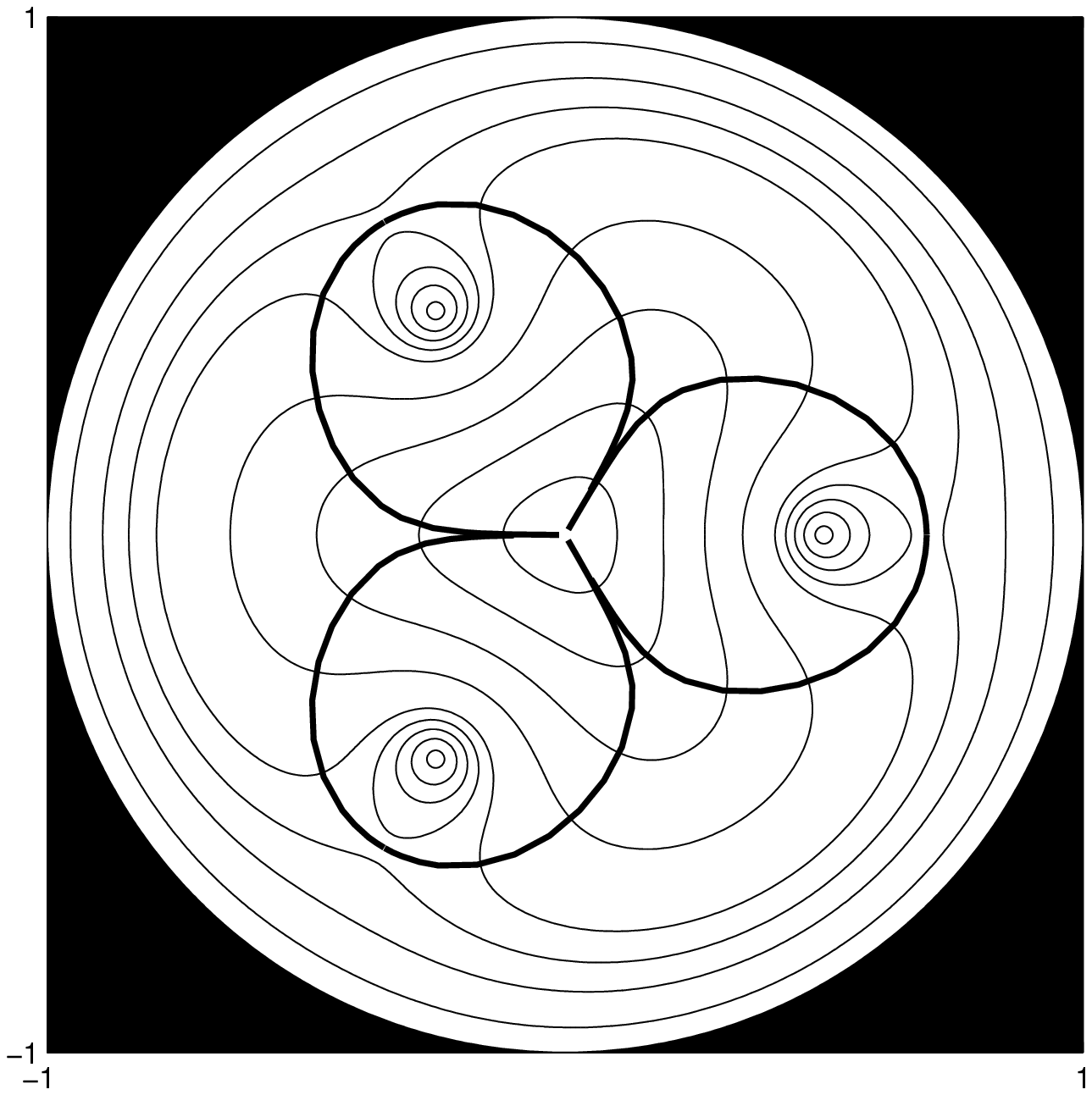,height=4.5cm,angle=0} } 
{\vspace*{.13in}} 
\caption{The case of $N=3$  vortices  for  a). $\phi_m (3)<\phi_e < \phi_M(3)$,
b).  $\phi_M (3)<\phi_e< \phi_c(3)$ and c).  $\phi_e > \phi_c(3)$.
The curve $\Gamma$ is shown. } 
\label{gamma3} 
\end{figure}

In that case, the central point is  a minimum for all values of $\phi_e$. 
In addition for low enough 
external flux there are $N$ saddle points inside the disk 
situated symmetrically at $\theta=2\pi(k+1/2)/N$.  
As the external flux is increased through  $\phi_m(N)$, where
\begin{equation} 
\phi_m (N) =N\frac{1-x^N}{1+x^N}. 
\label{pim} 
\end{equation} 
 $N$ pairs of saddle  
points on the boundary appear  
together with $N$ minima in the bulk (see Fig.~\ref{gamma3}a). 
The minima move towards the  
center along the line defined by $\theta=2\pi (k+1/2)/N$, $k=1,\ldots,N$,  
while two saddle points constituting 
the $k$-th pair move symmetrically  to the points $\theta=2\pi k/N$ and  
$\theta=2\pi (k+1)/N$  as the flux increases in the interval  
$\phi_m(N) <\phi_e<\phi_M(N)$.
 When $\phi_e=\phi_M(N)$, with
\begin{equation} 
\phi_M (N) =N\frac{1+x^N}{1-x^N}  
\label{piM} 
\end{equation} 
 pairs of   saddle points 
meet at  $\theta=2\pi k/N$ and for  larger values of the flux there are 
$N$ saddle points situated   between a vortex and the boundary.
 In this regime  
there are $N+1$ minima and $2N$ saddle points inside the disk as shown
in~Fig.~\ref{gamma3}b. 
 
At $\phi_e=\phi_c(N)$, where $\phi_c(N)$ is given by the 
formulae (\ref{rhoc}) and  
(\ref{phic}), each of the $N$ minima coalesce with each of
 the $N$ saddle points 
at $r=\rho_c^{1/N},\;\theta=2\pi (k+1/2)/N$, where $\rho_c$ is given by 
(\ref{rhoc}).   
For $\phi_e>\phi_c(N)$  there remains only one minimum  
in the center and $N$ saddle points. It can be shown
 that $\phi_c (N)$ is always 
greater than $\phi_M (N)$ for $N>2$.

Then, for $\phi_e=\phi_c(N)$ a topological  
transition occurs, when a minimum and a saddle point coalesce. This transition and the  
corresponding curve $\Gamma$ is depicted in Fig.~\ref{gamma3}b 
and \ref{gamma3}c. It is 
 seen that the curve $\Gamma$ undergoes itself a topological  
transition. For $\phi_e<\phi_c(N) $ it does not touch the center and has a  
topology of a ring, while for $\phi_e>\phi_c(N) $ it has a form 
of petals of a  
flower emanating from the center of the disk.

 \subsection{A dynamical interpretation of  the curve $\Gamma$}

\begin{figure} 
{\hspace*{-0.2cm} 
\psfig{figure=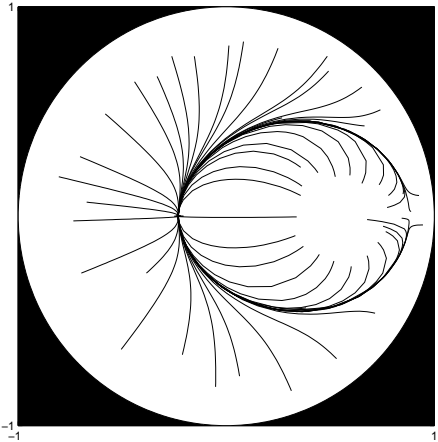,height=6cm,angle=0}} 
{\vspace*{.13in}} 
\caption{The curve $\Gamma$ as the limit cycle of the flow 
${\vec \nabla} h$. The parameters are the same as in Fig.~\ref{gamma1}a.} 
\label{limcy} 
\end{figure}

The curve $\Gamma$, defined above as being
 everywhere normal to the current flow,  has a simple interpretation
in terms  of an effective dynamical 
system, which gives an alternative method to compute it.
 Instead of the hamiltonian flow  
(\ref{hameq}), consider the lines of $\vec{\nabla} h$, obtained 
from those of $\vec{\jmath}$  
by a rotation of $90^0$. This flow is clearly non-Hamiltonian 
and possess fixed points which 
are not present in the hamiltonian dynamics, {\it e.g.} 
the positions of the vortices appear  
in this picture as a point-like sources, though a
 saddle remain a saddle in this  
transformation. Then, the curve $\Gamma$ represents a limit cycle of the  
flow $\vec{\nabla} h$ emanating from the saddle point. 
This is illustrated in Fig.~\ref{limcy} 
where it is clearly seen that the same curve as in 
 Fig.~\ref{gamma1}~a)  appears as a  
limit of the lines of $\vec{\nabla} h$.

\section {Conclusion}

 In this paper we have presented two different aspects of vortex patterns 
 in a  mesoscopic superconductor. 
  First we focused on  equilibrium 
 properties of a type II  superconductor in the London regime. 
 In this case, the Ginzburg-Landau equations reduce to a single linear 
 elliptic equation that can explicitly be solved in simple geometries. 
 We have derived closed expressions for the free energy and the magnetization of 
 a mesoscopic system which have been obtained  as limiting cases  
 of  a  general  calculation valid for a system of any size  
 \cite{bobel} and not from specific assumptions related to the 
 mesoscopic regime \cite{bb}. With  these formulae, we have been able to study 
 precisely the matching fields and to compute  
 the paramagnetic  Meissner effect, resulting from the existence 
 of metastable  states. 
  The second part of the paper has been devoted to the general and topological 
 study of the metastable states using a parallel with  
the theory of two dimensional dynamical systems. We showed that 
 a vortex pattern can be characterized by as set of topological integers, 
 the number of vortices being one of them. Any change in this set  
corresponds to a topological transition of the vortex pattern. 
 We calculated  the critical fields associated with these transitions 
 and we gave  a dual interpretation of these transitions. A  key concept 
 is the introduction of a special  curve $\Gamma$, that embodies 
 the main geometric features of a configuration of vortices. 
 This curve, which appears mathematically  as a limit cycle of the 
 system of currents generated by the vortex pattern, has a simple 
 physical meaning: it  separates paramagnetic from diamagnetic 
 domains.  
 
 We emphasize that this approach that focuses  on  topological 
 aspects  are generic and can be extended, at least in a qualitative 
 manner,  to  systems beyond the London regime and with different shapes. Here, we considered 
  a circular cylinder  to obtain  simple analytical expressions. 
 In recent  studies,  analytic expressions for the free energy  
 of a  cylinder  with an elliptic cross-section  were found  
  in the mesoscopic limit  \cite{meyers} 
 and,  more generally,   an electrostatic analogy was  
 developed  that enables to study mesoscopic superconductors 
 in the London regime with the help of conformal transformations 
 \cite{bdaumens}; using such tools,  it would be interesting to carry on  
 a quantitative study of topological phase transitions in arbitrary  domains. 
 
 In this work, only static configurations of vortices have been investigated. 
 A natural extension of the concepts introduced  here would be to consider the 
 dynamics of a vortex pattern, where  a vortex would move in the field 
 generated by the other vortices. In this context,  
 the study of the deformations of the curve 
 $\Gamma$ would certainly  shed some light on the 
 mechanism of vortex nucleation in a superconducting system.

\section{Acknowledgments} 
 
E.A and D.M.G. would like to thank Mark Mineev for fruitful and stimulating discussions. 
D.M.G. would like to thank SPhT  at Saclay for  hospitality.  
This research was supported in  part by  
the U.S.--Israel Binational Science Foundation (BSF), by the Minerva  
Center for Non-linear Physics of Complex Systems, by the Israel  
Science Foundation, by the Niedersachsen Ministry of Science  
(Germany) and by the Fund for Promotion of Research at the  
Technion.  
 
\begin{appendix}

\section{Magnetic induction in a cylinder \label{aplondon}} 
 
We present the solution of the London equation (\ref{eqh1}) for the magnetic induction 
\begin{equation} 
\nabla^2 h(\vec{r}) -2 h (\vec{r}) =-4\pi \sum_{k=1}^N \delta (\vec{r} - \vec{r}_k) 
\label{eqh1a} 
\end{equation} 
with the boundary condition $h(R,\theta)=h_e$.  
The polar coordinates are convenient for this geometry. 
The equation (\ref{eqh1a}) with a vanishing right hand side describes the  
Meissner regime in the absence of vortices.  
The corresponding  solution of the homogeneous equation is given by the  
$\theta$-independent function  
\begin{equation} 
h_M(r) = h_e \frac{I_0 (\sqrt{2}r)}{I_0(\sqrt{2}R)}.  
\end{equation} 
where the functions $I_n (x)$, $K_n (x)$ are modified Bessel functions  
of first and second kind of order $n$ which provide the regular and singular  
solutions respectively of (\ref{eqh1a}).   
The  solution of the inhomogeneous equation  (\ref{eqh1a}) for  $N$ vortices is written as a sum of  
the solution of the homogeneous equation and a particular solution: 
\begin{equation}  
h (r,\theta) =h_M (r)+ h_V (r,\theta)+h_{\bar{V}} (r,\theta)   
\label{h} 
\end{equation} 
where $h_V (r,\theta)$ is  the solution for $N$ vortices in the unbound domain  
\begin{equation}  
 h_V (r,\theta) = 2\sum_{k=1}^N K_0 (\sqrt{2}|\vec{r}-\vec{r}_k|) 
\label{eqhinf} 
\end{equation}  
and the term $h_{\bar{V}} (r,\theta)$ is introduced to take care of  the  
boundary conditions: 
\begin{equation} 
h_{\bar{V}} (R,\theta) =-h_V (R,\theta)= - 2\sum_{k=1}^NK_0 \left(\sqrt{2\left(R^2+r_k^2-2Rr_k\cos(\theta-\theta_k)\right)}\right) 
\label{bc1} 
\end{equation} 
On the other hand, the  field $h_{\bar{V}}$ can be written as a superposition  
of solutions of (\ref{eqh1a}) regular in the  interior of the boundary: 
\begin{equation} 
h_{\bar{V}} (R,\theta) = \sum_{n=-\infty}^{+\infty} c_n I_n (\sqrt{2} r) e^{in\theta} 
\label{super} 
\end{equation} 
where, since the field is real, the relation  $c_{-n} = c^*_n$  holds. 
Expanding the lhs of (\ref{bc1}) into Fourier series using standard identities of Bessel functions \cite{abrasteg} we get 
\begin{equation} 
K_0 \left(\sqrt{2\left[R^2+r_k^2-2 R r_k \cos(\theta-\theta_k)\right]}\right) = \sum_{n=-\infty}^{+\infty}  
K_n (\sqrt{2}R)I_n(\sqrt{2} r_k) e^{in(\theta-\theta_k)}. 
\label{expbessel} 
\end{equation} 
Equating the coefficients of the expansion of both sides of equation (\ref{bc1}) one obtains 
\begin{equation} 
c_n = \frac{K_n(\sqrt{2}R)I_n(\sqrt{2}r_k)}{I_n(\sqrt{2}R)}e^{-in\theta_k} 
\label{cn} 
\end{equation} 
from which the expression (\ref{hVbar}) follows immediately.

\section{Expression of the Gibbs energy  for a cylinder \label{apgibbs}}  
 
One takes advantage of the vector identity  
$ 
(\nabla\times \vec{h})\cdot(\nabla\times\vec{h}) =  
\nabla\cdot (\vec{h}\times\nabla\times\vec{h})+ 
\vec{h}\cdot(\nabla\times\nabla\times\vec{h}) 
$ 
to rewrite $(\nabla\times \vec{h})\cdot(\nabla\times\vec{h}) = \nabla\cdot (\vec{h}\times\nabla\times\vec{h})- 
\vec{h} \nabla^2\vec{h}$, since $\nabla\cdot\vec{h} = 0$.  
Then one uses the London equation (\ref{eqh1})  to rewrite the expression (\ref{freeenergy}) for the  
free energy as 
\begin{equation} 
{\cal F} = \pi\sum_{k=1}^N \int_\Omega d^2r\, h(\vec{r})\delta(\vec{r}-\vec{r}_k) + 
\frac{1}{4} \int_\Omega d^2r\, \nabla\cdot (\vec{h}\times\nabla\times\vec{h}) 
\label{free1} 
\end{equation} 
The singular part ${\cal F}_\infty$ is obtained from the contribution 
 of $h_V$ to the first integral. The regular  
contribution  to  the free energy can be written using  Stokes theorem: 
\begin{equation} 
{\cal F}-{\cal F}_\infty = \pi\sum_{k=1}^N h_M(\vec{r}_k) + \pi\sum_{k=1}^N h_{\bar{V}} (\vec{r}_k)+ 
\frac{R}{4} \int_0^{2\pi} d\theta\, h(R,\theta) h_r (R,\theta) 
\label{free1sing} 
\end{equation}   
where $h_r$ denotes the partial derivative of $h(r,\theta)$ with respect to $r$ and the fields  
$h_M$ and $h_{\bar{V}}$ are given by (\ref{hM}) and (\ref{hVbar}). In the last term $h(R,\theta)=h_e$  
can be taken out of the integration sign and the integral becomes 
\begin{eqnarray} 
&&2\pi  \frac{d h_M (R) }{d R} +  
2\pi \frac{\partial }{\partial R}\int_0^{2\pi} d\theta\,  
\left( h_V (R,\theta) + h_{\bar{V}} (R,\theta)\right) \nonumber \\ 
&=&  
2\sqrt{2} \pi h_e \frac{I_1(\sqrt{2}R)}{I_0(\sqrt{2}R)} - 
4\sqrt{2}\pi\sum_{k=1}^N \frac{I_0(\sqrt{2}r_k)}{I_0(\sqrt{2}R)}  
\left(I_0(\sqrt{2}R) K_1(\sqrt{2}R) + K_0(\sqrt{2}R)  I_1(\sqrt{2}R) \right) 
\label{integr} 
\end{eqnarray} 
To prove the last equality  the properties of the Bessel functions $I_0'(x)=I_1(x)$, $K_0'(x)=-K_1(x)$  
and the expansion (\ref{expbessel}) were used. Only $n=0$ term contributes to the angular integral. 
Using another property of Bessel functions   
$I_0(x) K_1(x) + K_0(x)  I_1(x)  =1/x$ 
and one obtains that: 
\begin{equation} 
\frac{h_e R}{4}\int_0^{2\pi} d\theta\, h_r (R,\theta) = 
\frac{\sqrt{2}\pi R h_e^2}{2} \frac{I_1(\sqrt{2}R)}{I_0(\sqrt{2}R)} -  
\pi\sum_{k=1}^N h_e \frac{I_0(\sqrt{2}r_k)}{I_0(\sqrt{2}R)} 
\label{hr} 
\end{equation} 
We notice  that the contribution from the Meissner field $h_M$ in (\ref{free1}) exactly   
cancels with the second term in the last equation. Hence, the regular part of free energy becomes: 
\begin{equation} 
{\cal F}-{\cal F}_\infty = \sqrt{2}\pi R  h_e^2\frac{I_1(\sqrt{2}R)}{I_0(\sqrt{2}R)} - 
2\pi \sum_{n=-\infty}^{+\infty}\sum_{j,k=1}^N  
\frac{K_n(\sqrt{2}R)I_n (\sqrt{2}r_j)I_n (\sqrt{2}r_k)}{I_n(\sqrt{2}R)}\cos n(\theta_j-\theta_k) 
\label{free2} 
\end{equation}   
 
The term containing the average magnetic induction over the cross section of the cylinder  
is evaluated along the same lines,  using the London equation (\ref{eqh1}): 
\begin{equation} 
h_e\int_{\Omega} h({\vec r}) d^2 r =2\pi Nh_e+  
\frac{h_e}{2}\int_{\Omega} \nabla\cdot\nabla h({\vec r}) d^2 r  =  
 \sqrt{2} \pi R h_e^2 { I_1( \sqrt{2}R) \over I_0 ( \sqrt{2}R)} +  
 2 \pi h_e \sum_{k = 1}^{N} \left( 1 - { I_0 ( \sqrt{2}r_k) \over I_0 ( \sqrt{2}R)}  \right) 
\label{average} 
\end{equation} 
Using  the expressions (\ref{free2}) and (\ref{average})  in the definition  
(\ref{gibbs}) of the Gibbs energy completes the derivation of (\ref{gibbse}). 
 
\section{Study of  the critical points \label{appc}}

In order to  study  critical points of the Hamiltonian system defined
 in section 5.1, one has to solve the equations  (\ref{dhdr})  and
(\ref{dhdtheta}) for a given number $N$ of vortices.

\subsection{Existence and location of   the critical points}

We first look for critical points which are neither on the boundary of the disk
 nor at the center, {\it i.e.} we concentrate on the generic position
  $0<r<1$. When  $0<r<1$,  one deduces that 
 the   equation (\ref{dhdtheta}) is satisfied  
if $\sin N\theta =0$, {\it i.e.} when $\theta=k \pi/N$ for  
$k=0,1,\dots , 2N-1$. These values of the angle
  can be subdivided into two groups  
according to the sign of $\cos N\theta$. 
We now discuss the solution of the equation (\ref{dhdr}) for the two cases  separately. 

 \subsubsection{Bulk critical points such that  $\cos N\theta=+1$.}
 
\begin{figure} 
{\hspace*{-0.2cm} 
\psfig{figure=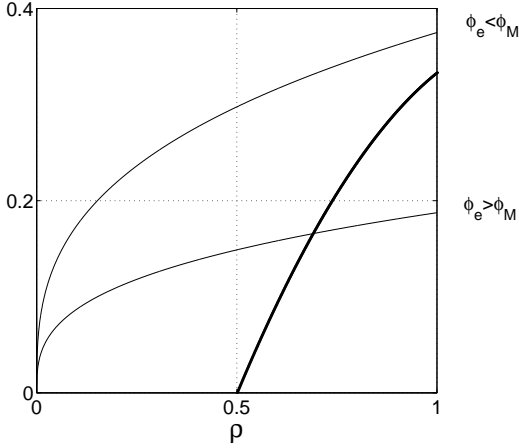,height=6cm,angle=0}} 
{\vspace*{.13in}} 
\caption{Graphical solution of the equation (\ref{case1rec}) for
 $N>2$ vortices. 
The position of the ring of vortices  is taken to be $x=0.5$.} 
\label{gcase1} 
\end{figure}

 In this case, we are looking for 'parallel'  critical points that lie in the
 same direction as the vortices. 
 And we shall prove that for any $N$, there  exists
 a critical field $\phi_M (N)$, defined as follows,
\begin{equation} 
\phi_M (N) = N (1+x^N)/(1-x^N) 
\label{phiM} 
\end{equation} 
such that when $\phi_e > \phi_M (N)$
there is always a critical  
point between each  vortex and the boundary, and no such point exists if  
$\phi_e <\phi_M (N)$. 
 \newline\newline  
 Using the new  variables $\xi=x^N$ and $\rho=r^N$ the equation 
(\ref{dhdr}) reduces to 
\begin{equation} 
\frac{\phi_e}{N}\rho^{\frac{2}{N}-1} = \frac{\xi^2-1}{(1-\xi\rho)(\xi-\rho)} 
\label{case1} 
\end{equation} 
Looking at this equation, it is clear that systems with $N=1$, $N=2$ or $N>2$
 should be  discussed separately.
 \begin{itemize}
 \item  For $N=1$, the plots of the left and right hand sides  of   
this equation are  
represented in Fig.~\ref{gcase1n1}. They cross for $\rho<1$ only if  the rhs 
evaluated at $r=1$ is less than  $\phi_e$, {\it i.e} for  
$\phi_e >\phi_M = (1+\xi)/(1-\xi) = (1+x)/(1-x)$. In this case the solution of (\ref{case1})  
can be shown to 
exist  using the continuity of the functions.  The position of the critical point  
corresponding to this solution satisfies $\xi<r<1$, {\it i.e} the point resides  between the  
vortex and the boundary. For $\phi_e <\phi_M$ there is no  solution. 
\begin{figure} 
{\hspace*{-0.2cm} 
\psfig{figure=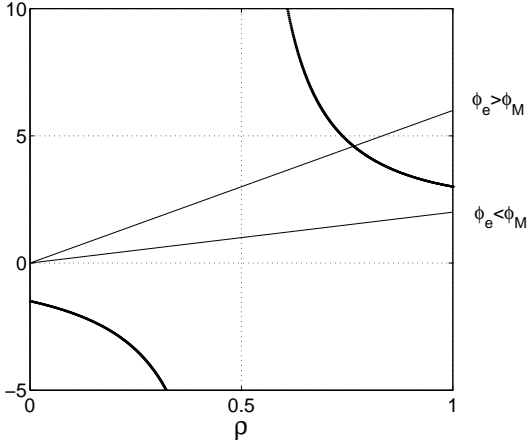,height=6cm,angle=0}} 
{\vspace*{.13in}} 
\caption{Graphical solution of the equation (\ref{case1}) for one single vortex ($N=1$). 
The position of the vortex is taken to be $x=0.5$.} 
\label{gcase1n1} 
 
\end{figure} 
 
 \item For $N\ge 2$, it is worthwhile to consider the reciprocal of (\ref{case1}) which  
rewrites as 
\begin{equation} 
\frac{N}{\phi_e}\rho^{1-\frac{2}{N}} =\frac{1}{\xi^2-1} (1-\xi\rho)(\xi-\rho) 
\label{case1rec} 
\end{equation} 
The right hand side is a convex parabola with zeros at $\xi<1$ and $1/\xi>1$.  
For $N=2$ the lhs is a constant function and it is clear that if  
$\phi_e >\phi_M = 2 (1+\xi)/(1-\xi) = 2 (1+x^2)/(1-x^2)$ 
there is always a solution for $\xi<\rho<1$, therefore there exists a critical point 
between the vortex and the boundary, while $\phi_e <\phi_M$ no solution exists. 
For $N>2$ the lhs is an increasing fractional power of $\rho$ (see Fig.~\ref{gcase1}). 
The equation (\ref{case1rec}) always has a solution 
between $\xi$ and $1$ if $\phi_e >\phi_M = N (1+\xi)/(1-\xi) = N (1+x^N)/(1-x^N)$ and  
no solution if $\phi_e <\phi_M$.  
 \end{itemize}

\subsubsection{Bulk critical points such that $\cos N\theta=-1$.}

 We  are  now looking for `anti-parallel'
 critical points that lie in a direction 
 bisecting the angle between two neighboring vortices. Here the
 discussion is more involved: 

\begin{itemize}

 \item when $N=1$ there appears a  critical field
 $\phi_m (1) = (1-x)/(1+x)$ such that:  when $\phi_e > \phi_m (1)$ there
 exists  a  critical point in the direction opposite to the vortex
 whereas when  $\phi_e<\phi_m (1)$  no such point  exists. 
  
 \item  when $N =  2,$  there are 
two critical fields  $\phi_m(2)$ and  $\phi_c(2)$
  such that: when $\phi_m(2)< \phi_e<\phi_c (2)$ 
 there are  exactly   two  anti-parallel  critical points
 and there are  no such points  when
 $\phi_e<\phi_m (2)$ or  $\phi_e>\phi_c (2)$. 
 
\item  when $N > 2,$ one has again  
two critical fields  $\phi_m(N)$ and  $\phi_c(N)$ such that:
 when 
 $\phi_e<\phi_m (N)$ there exist   $N$ anti-parallel  critical points,
 when  $\phi_m(N)< \phi_e<\phi_c (N)$, there are $2N$ such points,
 and  when  $\phi_e>\phi_c (N)$, there are no anti-parallel  critical points.
\end{itemize}

 We now prove these assertions and calculate the corresponding
 critical fields.

\begin{figure} 
{\hspace*{-0.2cm} 
\psfig{figure=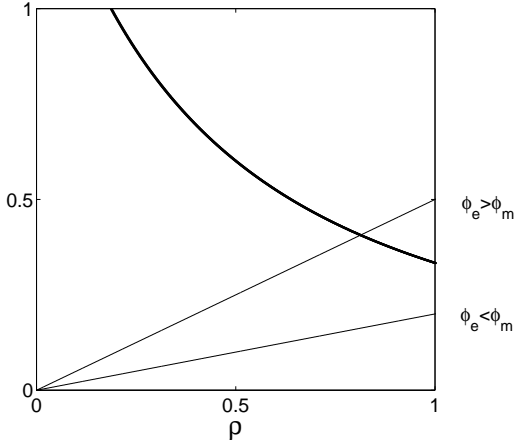,height=6cm,angle=0}} 
{\vspace*{.13in}} 
\caption{Graphical solution of the equation (\ref{case2}) for one single vortex ($N=1$). 
The position of the vortex is taken to be $x=0.5$.} 
\label{gcase2n1} 
\end{figure}

  Consider first  that  $N=1$;  then 
the equation (\ref{dhdr}) in terms of the variables  
$\rho$ and $\xi$ becomes 
\begin{equation} 
 {\phi_e}\rho = \frac{1-\xi^2}{(1+\xi\rho)(\xi+\rho)} 
\label{case2} 
\end{equation} 
 And  a solution exists in the interval
 $0<\rho<1$ if the value of the lhs at  
$\rho=1$ exceeds the value of the rhs at the same point, as shown in  
Fig.~\ref{gcase2n1}. 
This happens for $\phi_e>\phi_m (1) = (1-x)/(1+x)$; 
 
 Now, when  $N\ge2$,  the equation (\ref{dhdr}) can be rewritten as  
\begin{equation} 
\frac{N}{\phi_e}\rho^{1-\frac{2}{N}} = \frac{1}{1-\xi^2}(1+\xi\rho)(\xi+\rho) 
\label{case2rec} 
\end{equation} 

\begin{itemize}
 \item  $N=2$: this case is  illustrated in Fig.~\ref{gcase2n2}. For  
$\phi_e <\phi_m (2)  = 2(1-\xi)/(1+\xi) = 2(1-x^2)/(1+x^2)$ there
 is no solution, for 
$\phi_e=\phi_m (2)$ one solution appears at $\rho=1$ and 
it persists for the range of  
the flux  
$\phi_m(2) <\phi_e <\phi_c (2)=2 (1-\xi^2)/\xi=2(1-x^4)/x^4$.   
For higher flux, $\phi_e>\phi_c (2)$, this solution disappears at $\rho=0$.
 
\begin{figure} 
{\hspace*{-0.2cm} 
\psfig{figure=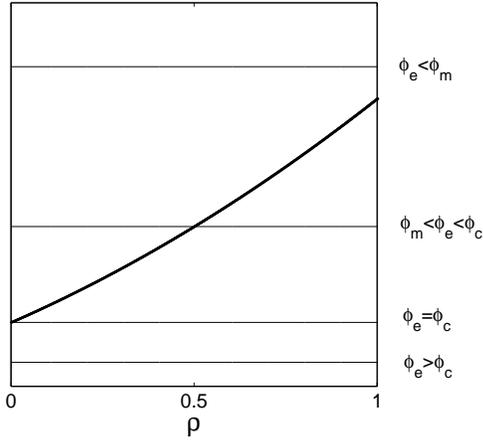,height=6cm,angle=0}} 
{\vspace*{.13in}} 
\caption{Graphical solution of the equation (\ref{case2rec}) for $N=2$ vortices. 
The position of each vortex is taken to be  $x=0.5$.} 
\label{gcase2n2} 
\end{figure} 

 \item The case $N>2$ is the  most interesting:
 Fig.~\ref{gcase2} shows the behavior  
of both sides of 
the equation (\ref{case2rec}). Comparing the value of the lhs at $\rho=1$ with that  
of the rhs and using the fact that $\rho^{1-\frac{2}{N}}$ has a negative curvature,  
while that of the rhs is always positive,  we can state that for  
$\phi_e <\phi_m (N)$, where 
\begin{equation} 
\phi_m (N) = N(1-\xi)/(1+\xi) = N(1-x^N)/(1+x^N) 
\label{phim} 
\end{equation} 
 there is only one solution. Another  
solution appears at $\rho=1$ for $\phi_e =\phi_m (N)$ and these two solution  
move towards each other when the flux is increased in the interval  
$\phi_m (N) <\phi_e<\phi_c (N)$, whereas at the  
critical value of the flux $\phi_e=\phi_c (N)$  these two solutions coalesce. For    
$\phi_e>\phi_c (N)$, the equation (\ref{case2rec}) has no solutions in the  
interval $0<\rho<1$. 
\begin{figure} 
{\hspace*{-0.2cm} 
\psfig{figure=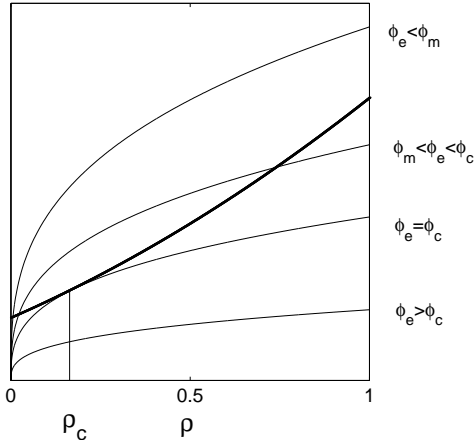,height=6cm,angle=0}} 
{\vspace*{.13in}} 
\caption{Graphical solution of the equation (\ref{case2rec}) for $N>2$ vortices. 
The position of each vortex is taken to be  $x=0.5$.} 
\label{gcase2} 
\end{figure} 
 
\end{itemize}

\noindent\underline{Calculation of  the critical flux $\phi_c$:}
    
For $\phi_e (N)=\phi_c $ the function  
$f_{lhs}(\rho) = \frac{N}{\phi_e}\rho^{1-\frac{2}{N}}$ and 
$f_{rhs} (\rho) =(1+\xi\rho)(\xi+\rho)/(1-\xi^2)$ are tangent at some point  
$\rho=\rho_c$ as shown in Fig.~\ref{gcase2}. Equating the value of these functions  
and the value of their derivatives at $\rho=\rho_c$ we obtain the following equations: 
\begin{equation} 
\left\{\begin{array}{ccc} 
\phi_c (\rho_c+\xi)(1+\xi\rho_c) &=& N(1-\xi^2)\rho_c^{1-\frac{2}{N}}\\ 
\phi_c (2\xi\rho_c+\xi^2+1) &=& (N-2)(1-\xi^2)\rho_c^{-\frac{2}{N}} 
\end{array}\right. . 
\label{tangeq} 
\end{equation} 
Dividing the first equation by the second  leads to the quadratic equation for $\rho_c$: 
\begin{equation} 
(N+2)\xi\rho_c^2+2(\xi^2+1)\rho_c-(N-2)\xi=0 
\label{quadratic} 
\end{equation} 
which has the unique positive solution 
\begin{equation} 
\rho_c=\frac{-(1+\xi^2)+\sqrt{(1-\xi^2)^2+N^2\xi^2}}{(N+2)\xi} 
\label{rhoc} 
\end{equation} 
The critical flux is obtained by substituting the value of $\rho_c$ into the first 
equation of (\ref{tangeq}): 
\begin{equation} 
\phi_c= \frac{N(1-\xi^2)\rho_c^{1-\frac{2}{N}}}{(\rho_c+\xi)(1+\xi\rho_c)} 
\label{phic} 
\end{equation} 
We notice that for  $N=2$ the formula (\ref{rhoc}) yields the correct value $\rho_c=0$. 
Although topologically the cases $N=2$ and $N>2$ are different,  the critical flux 
$\phi_c(2)=(1-\xi^2)/\xi$, as we shall see later, comes out correctly.

\subsubsection{Central and boundary critical points}
 
So far we did not discuss the two cases, $r=0$ and $r=1$. For $N=1$ the center, $r=0$,  
of the disk  does not satisfy the equations (\ref{dhdr}),(\ref{dhdtheta}) and  
therefore is not a  
critical point. For $N>1$ the value  $r=0$ is always a solution of the equations  
(\ref{dhdr}),(\ref{dhdtheta}). For the points on the boundary, $r=1$,  the equation  
(\ref{dhdtheta}) is always satisfied and substituting the value $r=1$  
into (\ref{dhdr}) we find that  
there are $2N$ critical  points on the boundary with an angle given by the  
following relation: 
\begin{equation} 
\cos N\theta = \frac{1}{2x^N}\left[ 1+x^{2N}-\frac{N(1-x^{2N})}{\phi_e}\right] 
\label{cosntheta} 
\end{equation} 
which is meaningful only for values of the flux between $\phi_m (N)$ and  
$\phi_M (N)$. 

\subsection{The nature of the critical points}
 
We now discuss the topological nature of the critical points and classify them 
by linearizing the  
stability  exponents  near each of these points.  
The problem is then reduced to the study of the
 Hessian matrix of the second derivatives  
\begin{equation} 
H=\left(\begin{array}{cc} 
\partial^2_r h & \partial_r\partial_\theta h \\ 
\partial_\theta\partial_r h & \partial^2_\theta h 
\end{array}\right) 
\label{hessian} 
\end{equation} 
Since at the critical points inside
 the disk we have $\sin N\theta =0$ the mixed  
derivative vanish 
and the Hessian matrix is diagonal in polar coordinates. The second derivative with  
respect to $\theta$ is given at the critical points by 
\begin{equation} 
\partial^2_\theta h = -2N^2 \rho \frac{(1-\rho^2)(1-\xi^2)}{(1-\xi\rho)^2(\xi-\rho)^2}<0 
\label{d2hrcase1} 
\end{equation} 
 for  $\cos N\theta = +1$ and by 
\begin{equation} 
\partial^2_\theta h = +2N^2 \rho \frac{(1-\rho^2)(1-\xi^2)}{(1+\xi\rho)^2(\xi+\rho)^2}>0 
\label{d2hthcase1} 
\end{equation} 
for $\cos N\theta = -1$. 
The value of the second derivative with respect to $r$ can be obtained by looking at the  
way the first derivative $\partial_r h$
 changes sign near each solution of the equation  
(\ref{dhdr}). We shall now discuss separately the nature
 of the critical points in the  'parallel', 'antiparallel', 'central'
 and 'boundary' cases.

\subsubsection{Bulk critical points with $\cos N\theta=+1$.} 

 For $\phi_e>\phi_M$, there is  
only one  critical point at $\rho_1$ in the interval $\xi<r<1$.  
The results for the second derivatives at this point are
 summarized in the following  
table:  
 
\vspace{10pt} 
\centerline{ 
\begin{tabular}{|c|c|c|} 
\hline 
                & $0<\rho<\rho_1$       &   $\rho_1<\rho<1$     \\       
\hline 
$\partial_r h$  & --                    & +                     \\ 
\hline 
$\partial^2_r h$&               \multicolumn{2}{|c|}{ + }               \\ 
\hline 
$\partial^2_\theta h$   &       \multicolumn{2}{|c|}{ -- }              \\ 
\hline\hline 
Crit. point     &               \multicolumn{2}{|c|}{ Saddle}           \\ 
\hline 
\end{tabular}   } 
\vspace{10pt} 
 
\subsubsection{Bulk critical points with  $\cos N\theta = -1$.} 

  The following tables  
 provide a classification of critical points in a variety of regimes. 
 
\begin{description} 
\item[\underline{$N=1$}]  
         
        \begin{description} 
        \item[$\phi_e<\phi_m$] No critical points. 
        \item[$\phi_e>\phi_m$] One critical point at $\rho_1$ 
 
\vspace{10pt} 
\centerline{ 
\begin{tabular}{|c|c|c|} 
\hline 
                & $0<\rho<\rho_1$       &   $\rho_1<\rho<1$     \\       
\hline 
$\partial_r h$  & --                    & +                     \\ 
\hline 
$\partial^2_r h$&               \multicolumn{2}{|c|}{ + }               \\ 
\hline 
$\partial^2_\theta h$   &       \multicolumn{2}{|c|}{ + }               \\ 
\hline\hline 
Crit. point     &               \multicolumn{2}{|c|}{Minimum }          \\ 
\hline 
\end{tabular}   } 
\vspace{10pt} 
 
        \end{description} 
\item[\underline{$N=2$}] 
         
        \begin{description} 
        \item[$\phi_e<\phi_m$] No critical points. 
        \item[$\phi_m<\phi_e<\phi_c$] One critical point at $\rho_1$ 
 
\vspace{10pt} 
\centerline{ 
\begin{tabular}{|c|c|c|} 
\hline 
                & $0<\rho<\rho_1$       &   $\rho_1<\rho<1$     \\       
\hline 
$\partial_r h$  & --                    & +                     \\ 
\hline 
$\partial^2_r h$&               \multicolumn{2}{|c|}{ + }               \\ 
\hline 
$\partial^2_\theta h$   &       \multicolumn{2}{|c|}{ + }               \\ 
\hline\hline 
Crit. point     &               \multicolumn{2}{|c|}{Minimum }          \\ 
\hline 
\end{tabular}   } 
\vspace{10pt} 
 
        \item[$\phi_e>\phi_c$] No critical points. 
        \end{description} 
\item[\underline{$N>2$}] 
 
        \begin{description} 
        \item[$\phi_e<\phi_m$] One critical point at $\rho_1$  
 
\vspace{10pt} 
\centerline{ 
\begin{tabular}{|c|c|c|} 
\hline 
                & $0<\rho<\rho_1$       &   $\rho_1<\rho<1$     \\       
\hline 
$\partial_r h$  & +                     & --                    \\ 
\hline 
$\partial^2_r h$&               \multicolumn{2}{|c|}{ -- }              \\ 
\hline 
$\partial^2_\theta h$   &       \multicolumn{2}{|c|}{ + }               \\ 
\hline\hline 
Crit. point     &               \multicolumn{2}{|c|}{Saddle }           \\ 
\hline 
\end{tabular}   } 
\vspace{10pt} 
 
        \item[$\phi_m<\phi_e<\phi_c$] Two critical points at $\rho_1<\rho_2$ 
 
\vspace{10pt} 
\centerline{ 
\begin{tabular}{|c|c|c|c|c|c|c|} 
\hline 
  & \multicolumn{2}{|c|}{$0<\rho<\rho_1$}  &   \multicolumn{2}{|c|}{$\rho_1<\rho<\rho_2$} &  
\multicolumn{2}{|c|}{$\rho_2<\rho<1$}\\  
\hline 
$\partial_r h$  & \multicolumn{2}{|c|}{+}& \multicolumn{2}{|c|}{--} & \multicolumn{2}{|c|}{+}\\ 
\hline 
$\partial^2_r h$ & \multicolumn{3}{|c|}{ -- } & \multicolumn{3}{|c|}{ + }               \\ 
\hline 
$\partial^2_\theta h$   &\multicolumn{3}{|c|}{ + } & \multicolumn{3}{|c|}{ + }          \\ 
\hline\hline 
Crit. point     &\multicolumn{3}{|c|}{\ \ \ \ \ \ \ \ Saddle\ \ \ \ \ \ \ \  } & \multicolumn{3}{|c|}{Minimum }         \\ 
\hline 
\end{tabular}   } 
\vspace{10pt} 
 
        \item[$\phi_e>\phi_c$] No critical points. 
        \end{description} 
\end{description} 

 \subsubsection{Central and boundary critical points.}

Consider now the point $r=0$. Expanding the field $h(r,\theta)$ given by (\ref{hcirc}) 
for small $r$ we obtain 
\begin{equation} 
h(r,\theta)\approx -(\phi_e-2N\ln x )+\phi_e r^2 +2\cos (N\theta)\;\left(\frac{1-\xi^2}{\xi}\right)\; r^N 
\label{hsmallr} 
\end{equation} 
As we mentioned before, for $N=1$ the point $r=0$ is not a critical point  
(due to the presence of a  term linear in $r$ ). For $N=2$,  two last terms are  
quadratic in $r$  
and the character of the critical point depends on the value of the flux $\phi_e$. For  
$\phi_e<\phi_c=(1-\xi^2)/\xi$ the center, $r=0$,  is a saddle, while for  
$\phi_e>\phi_c$ it turns to be to a minimum. This is to be compared with the prediction
of formula \ref{phic}. It is interesting to notice  
that this happens when the  
minimum located along the line $\cos 2\theta =-1$ disappears at the center. For $N>2$, the  
last term in the equation (\ref{hsmallr}) is negligible and the center is always a  
minimum.   
 
When $\phi_m<\phi_e<\phi_M$ we must consider the critical points on the boundary. 
Since for $r=1$ the condition $\partial_\theta h =0$ is satisfied for all $\theta$,  
we obtain $\partial^2_\theta h = 0$ and  
$\partial_r\partial_\theta h \neq 0$ if $\cos^2 N\theta\neq 1$, which is always satisfied 
for $\phi_e$ in the interval between $\phi_m$ and $\phi_M$. 
 Thus the determinant of the Hessian matrix  
(\ref{hessian}), given by $\mbox{det}H = -(\partial_r\partial_\theta h)^2$, is negative  
and the eigenvalues are of opposite sign. We conclude 
that the critical points at the boundary are saddles.

\end{appendix}


\begin{thebibliography}{article}
\bibitem{rem} It should be noted here that  
in order to obtain this equation we must be in the London limit and impose  
$|\psi|^2 = 1$ since starting from (\ref{adgl2}), we have $ - \triangle h = |\psi {|^2} 
 ({\vec \nabla} \times {\vec \nabla} \chi - h)$ and where  
$({\vec \nabla} \times {\vec \nabla} \chi)_{\hat z} = 2 \pi \delta_2 (\vec r)$, we must in principle take $|\psi|^2 = 0$.
\end{thebibliography}
\end{document}